# Single-Photon Sensitive Optoelectronic Fibres for Distributed Nuclear Radiation Detection in Textile Fabrics

Nikhil Gupta[1,2], Hang Qi[3], Julian Kahlbow[3], Igor Korover[3], Areg Danagoulian[4], Or Hen[3] & Yoel Fink[1,2,5]

[1] Research Laboratory of Electronics, Massachusetts Institute of Technology, Cambridge, MA 02139, USA

[2] Department of Materials Science and Engineering, Massachusetts Institute of Technology, Cambridge, MA 02139, USA

[3] Department of Physics, Massachusetts Institute of Technology, Cambridge, MA 02139, USA

[4] Department of Nuclear Science and Engineering, Massachusetts Institute of Technology, Cambridge, MA 02139, USA

[5] Corresponding Author

# Abstract

Nuclear radiation detectors play a key role in applications spanning nuclear and particle physics, nuclear engineering, security, and medicine. With the expanded global interest in nuclear power, discreet, inconspicuous, and readily deployable nuclear detection capabilities are increasingly important. However, conventional dosimeters are often rigid, bulky, or lack spatial resolution, limiting their use for mobile, conformal, or large-area distributed mapping of dynamic fields. Here, we present flexible, radiation-sensitive optoelectronic fibres with up to 50% elasticity for real-time gamma dosimetry. Silicon photomultipliers are thermally drawn into the core of fibres composed of a scintillator waveguide, enabling electronic-photonic integration and detection of scintillation light with single-photon resolution. By modifying the device aspect ratio, we ensure the axially-facing orientation of the detection plane for efficient optical coupling. Two fibre architectures are developed: one with an annular plastic scintillating cladding, and the other with a liquid scintillator core in an elastomeric fibre. We show that these fibres are sensitive to localized nuclear radiation exposure from collimated 0.5 µCi Sr-90 β-sources and 10 µCi Cs-137 and Co-60 γ-sources, with extended responsivity measured over 30 cm, and estimated lower detection limits approaching near-background radiation levels (~14-41 nSv/hr). Co-locating the scintillator and detectors in the fibre addresses two perennial challenges associated with fibre-based systems: avoiding length limitations driven by optical losses and enabling a greater collection cone through capture of transient non-guided modes. We further enhance radiation sensitivity and mechanical robustness by covering the fibres with a tungsten-merino wool composite braid, enabling us to machine-weave them into fabrics alongside common textile yarns. The tungsten wires function as a gamma-electron converter, increasing the detection efficiency of the assembly by ~20%. Distributed woven arrays of fibres formed in this way present an opportunity to create large-area, conformal fabrics capable of real-time dosimetry of gamma radiation fields with high spatial resolution.

# Introduction

The application landscape for nuclear radiation detection continues to broaden, driven by critical needs across medical diagnostics, nuclear energy, national security, high-energy and nuclear physics, and materials analysis.[1–5] As detection systems are increasingly deployed in mobile, discreet, or constrained form factors,[6,7] the demand is growing for lightweight, real-time, spatially resolved technologies that can unobtrusively monitor dynamic radiation fields over large areas. In particular, there is increasing interest in distributed gamma-radiation dosimetry in low-to-moderate dose environments, such as human-occupied spaces, where conformal sensing platforms could enable continuous personal dosimetry and environmental monitoring on complex surfaces. In these settings, the primary objectives are rapid, high sensitivity dose-rate quantification using mechanically flexible distributed platforms, rather than discrimination between particle species or extreme radiation hardness. However, existing radiation detection technologies remain fundamentally limited in their ability to meet these requirements. Passive dosimeters, such as thermoluminescent or optically stimulated luminescent detectors, can record accumulated dose but require post-exposure laboratory processing, precluding their use in real-time applications.[8] Active radiation detectors, such as ionization chambers, silicon diode detectors, and Geiger-Müller counters, offer continuous monitoring but can be rigid, bulky, and power-hungry.[9–15] Notably, the majority of these common systems function as discrete or centralized point sensors, lacking the distributed architecture necessary to resolve spatial gradients across extended or nonplanar geometries.[16]

Scintillating fibre-coupled detectors offer improved spatial resolution by using arrays of scintillating fibres optically coupled to external photodetectors, such as photomultiplier tubes or silicon photomultipliers (SiPMs), allowing for position-sensitive radiation measurements.[17–23] In many implementations, the photodetectors are intentionally located outside the radiation field and externally coupled to optical fibres in order to protect sensitive electronics from radiation-induced damage. While this architecture is well-suited for high-fluence environments, the reliance on an external interface between the detector and fibre arrays often requires rigid support structures, large, centralized optical readouts, or relatively fragile configurations to ensure consistent optical coupling,[17–23] making them ill-suited for deployment on flexible or irregular surfaces. Additionally, their external coupling requirements, susceptibility to optical attenuation over extended fibre lengths, and signal routing constraints can limit scalable distribution across large surfaces, mechanical robustness in field applications, and ability to be integrated in functional fabrics and other media.

To address these limitations, there is a need for a new class of scalable nuclear radiation detectors that are flexible, lightweight, and capable of distributed, high-resolution sensing in real time. Textiles present a promising foundation since fabrics naturally offer large-area, conformal coverage of complex surfaces - including the human body - enabling new modes of unobtrusive and spatially extensive radiation monitoring.[24,25] Recent advances in the development of thermally drawn, device-containing fibres have led to their use in sensing applications across multiple physical domains, including thermal, optical, mechanical, acoustic, and chemical modalities.[26–32] Building on this foundation, textile-ready fibre-based nuclear radiation detectors have the potential to transform passive substrates into adaptive sensing systems, presenting new capabilities for continuous environmental awareness and personal dosimetry.

In this work, we present a thermally drawn, flexible, optoelectronic fibre-based radiation detector designed for real-time gamma dosimetry in conformal textile formats. This integrated, millimetre-scale electronic-photonic fibre detector is composed of a scintillating core material surrounded by an index-guiding cladding, with axially aligned silicon photomultipliers embedded directly within the fibre during the thermal draw. Upon interaction with an ionizing particle (e.g. gamma, neutron, or electron), the scintillating core emits optical photons, which are then guided by the fibre's waveguide structure toward the axially facing in-fibre SiPMs. The embedded devices perform in-fibre optoelectronic conversion, generating photon-number-resolving electrical signals across electrodes spanning the fibre's length. To characterize the detector's response in a stepwise manner, we progress from direct optical excitation with a pulsed laser, to direct energy deposition by beta particles, and finally to indirect excitation via gamma interactions producing secondary electrons. By localizing both light generation and periodic axial-facing detectors within the fibre, the system mitigates signal attenuation challenges common to externally coupled configurations. The modular design allows for photodetectors to be distributed along the fibre's length, enabling sensitivity to both beta and gamma radiation sources over extended distances, with estimated gamma detection limits approaching ~14-41 nSv/hr. A structure composed of high atomic number (high-Z) metallic wires interwoven with merino wool yarns is braided directly onto the fibre detector. By incorporating thin tungsten wires, which function as gamma-electron converters, the braid serves to increase detection sensitivity while maintaining flexibility and textile compatibility. The resulting composite is elastic, robust, and suitable for direct insertion into woven fabrics, laying the foundation for large-area, conformal gamma dosimetry systems with real-time, spatially-resolved capabilities.

## Fibre Fabrication and Architecture

### Hydrodynamic Stabilization for Radial-to-Axial Conversion

SiPMs allow for single-photon sensitivity with high detection efficiency in compact microelectronic devices.[33,34] Here, we use a 1600-microcell SiPM (AFBR-S4K11C0125B; Broadcom) with a 1 x 1 mm$^2$ active area and overall dimensions of 1.315 x 1.315 x 0.595 mm$^3$. Similar to other semiconductor photodetectors, SiPMs have a high-aspect ratio, planar geometry. To maximize coupling efficiency, the active surface of the SiPM should be oriented axially, i.e. perpendicular to the axis of the fibre, to facilitate efficient coupling with guided optical modes that originate in the scintillator. However, the extensional viscous forces imposed by the polymer fluid flow field during the thermal draw process tend to orient particulates in a manner that reduces drag and minimizes their hydrodynamic cross section.[35,36] This results in the active surface of the device facing radially as opposed to towards the fibre axis – the orientation that would be desirable.[37] To address this challenge, we introduce a method to orient the SiPM in the axial-facing direction **(Figure S1)**. To hydrodynamically stabilize the device against the rotational torque normally imposed by the draw process, we mould a high-temperature epoxy domain on the pad-side of the device that makes the axial dimension double the planar dimensions. This modified device geometry inverts the energetic preference through the thermal draw process, instead making the axial-facing configuration more stable than the radial-facing one **(Figure S2)**. Once stabilized, linear electrical connections are established between multiple SiPM devices through two 100-µm insulated copper wires attached to the anode and cathode of each device. The presence of a thin polyurethane insulating layer on the copper wires is crucial to maintain electrical isolation throughout the draw process. If the electrodes are shorted during the draw, exposure to draw temperatures can degrade the SiPM's performance and lead to increased dark count noise.

### Thermal Draw Process

We pursue two different fibre architectures: one based on a solid polymer scintillator, and the other using a liquid scintillator core. Each provides unique advantages; the use of a plastic scintillator allows for the realization of an all-solid-state fibre with incorporation of the active scintillating material during the draw process itself, while the use of a liquid scintillator allows for the mechanical properties of the fibre to be unconstrained by the scintillating material, enabling the realization of highly elastic fibres. In each case, the particular scintillator materials used for each fibre are chosen to have a wavelength of maximum emission between 420-430 nm to match the wavelength of peak photodetection efficiency of the SiPM. Additionally, because these materials are optimized to selectively produce fluorescent scintillation photons, the contribution of Cherenkov emission in these types of organic scintillators have been shown to be relatively minor and thus should not significantly affect the detected optical signal in the fibres [REF]. Furthermore, filtering or distinguishing the limited contribution of Cherenkov radiation is unnecessary for gamma dosimetry, where dose-rate measurements rely on total detected optical signal rather than particle-specific identification.

*Plastic Scintillator Fibre*

The fibre draw process begins with the fabrication of a macroscopic preform consisting of a hollow-core polyvinyltoluene (PVT) scintillator surrounded by a polymethylmethacrylate (PMMA) cladding in a rectangular geometry **(Figure S3)**. The PVT scintillator has a compatible softening temperature to PMMA, allowing for co-flow of the materials at high viscosity during the thermal draw process. In addition to acting as an index-guiding cladding, the outer PMMA domain redistributes the stress of the draw process from the scintillator to the cladding. This provides a viscous flow boundary condition that regulates the deformation of the inner PVT domain, which is prone to failure when drawn alone due to the significant elastic component of its rheological response **(Figure S4)**.[38] During the draw process, the axially-connected devices are fed into the large, hollow core of the preform **(Figure 1a)**. Because the maximum furnace temperature is 265 °C and the device passes through this region only briefly, the SiPM temperature remains below reflow soldering conditions. As a result, the SiPM withstands the thermal stresses of the draw with no observed degradation in performance over more than 10 device integrations, provided that the device electrodes remain insulated from one another during the draw. Due to the laminar co-flow of the preform materials, the fibre retains a rectangular annular waveguide structure of PVT cladded by PMMA, with the axial-facing SiPM embedded in the air core of the fibre **(Figure 1b-d)**. Because the SiPM faces the air core of the fibre, where optical modes guided in the annular PVT waveguide are evanescent, a local optical coupler is introduced post-draw directly adjacent to the SiPM **(Figure 1b)**. To do so, a 5-10 mm length of liquid photopolymer was injected next to the optically-active side of the device, and then UV-cured to form a solid optical coupling domain. The cured material is highly transparent and nearly index matched to the PVT scintillator at the wavelength of maximum emission, providing a nearly invisible optical boundary with minimal reflection or absorption. Simulations of a fibre without wires show that the coupling efficiency increases with the coupler length up to 5 mm, at which point the efficiency plateaus **(Figure S5)**. However, in practice, scattering from the wires in the core of the fibre will introduce greater optical losses for coupler lengths exceeding 5 mm. Additionally, variations in the SiPM epoxy moulding process and polymer flow during the draw can introduce misalignment of the SiPM relative to the fibre axis. Optical microscopy measurements across 5 embedded devices indicate a mean angular deviation from the fibre axis of 4.8° ± 2.7°. Although imperfect alignment can reduce the coupling efficiency, ray-tracing simulations show that even a 15° deviation from axial alignment results in less than a 10% reduction in optical coupling efficiency **(Figure S5)**.

*Liquid Scintillator Fibre*

To fabricate the elastic liquid fibre, axial-facing SiPMs were thermally drawn into a hollow core elastomeric cyclic olefin copolymer (ECOC) fibre by adapting a draw process described in previous work **(Figure 1e)**.[39] After the draw, the mechanical guide wires are removed from the optically-active side of the fibre by pinching the deformable elastomeric cladding at the device interface while pulling the wires from the fibre end, concentrating stress to fracture the wires near the device for full extraction. This procedure was performed on over 5 fibres without any observed damage to the embedded SiPM, as the low breakage force (~2 N) of the thin, 100 μm copper wires imposes relatively minimal mechanical stress on the device during removal. Then, the entire air channel length of the optically-active side of the fibre is filled by pumping in a high flash point liquid scintillator with low chemical toxicity (EJ-309). This process creates a cylindrical core-cladding structure for index-guiding light through the liquid scintillator core, with the photodetecting side of the SiPM directly interfacing with the liquid scintillator **(Figure 1f-h)**. Because in this case the wires would be within the scintillating waveguide itself, the removal of the mechanical wires is crucial to help reduce scattering of optical modes propagating through the scintillator core.

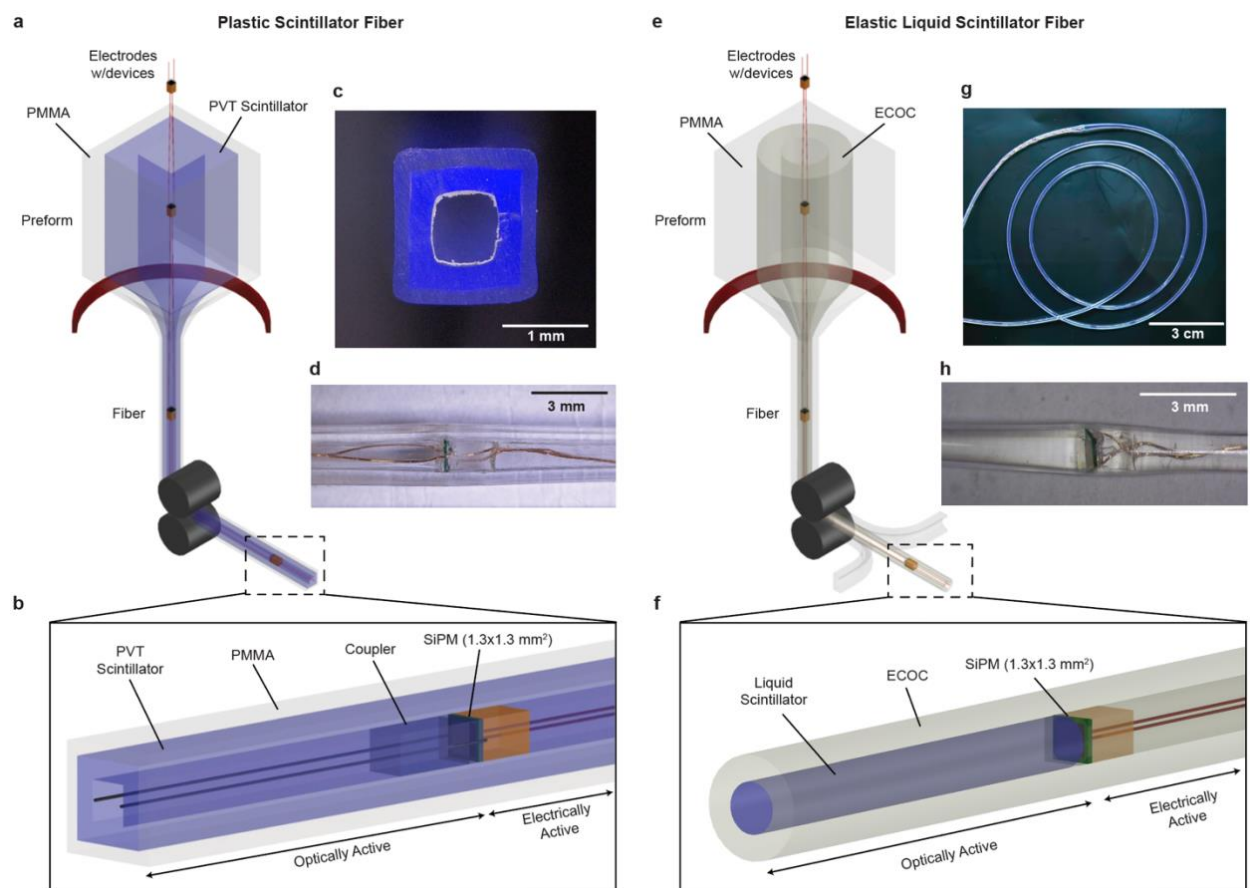

**Figure 1: Fibre Fabrication and Architectures** | **a.** Schematic of the thermal draw process used to fabricate the plastic scintillator fibre. **b.** Structure of the plastic scintillator fibre, consisting of a SiPM embedded in the air core of a rectangular fibre composed of an annulus of PVT scintillator cladded by PMMA. The optical coupler directly adjacent to the SiPM is injected into the fibre post-draw. **c.** Image of the cross section of a drawn plastic scintillator fibre. **d.** Image of a drawn plastic scintillator fibre with a SiPM embedded in its core. **e.** Schematic of the thermal draw process used to fabricate the liquid scintillator fibre. **f.** Structure of the liquid scintillator fibre, consisting of a SiPM embedded in the air core of a cylindrical fibre composed of ECOC. The mechanical guide wires are removed after the draw process, and the liquid scintillator is injected into the air core of the fibre. **g.** Image of a drawn liquid scintillator fibre, showing the contrast between the optically-active length, which scintillates blue, and the electrically-active length containing the wires. **h.** Image of a liquid scintillator fibre with a SiPM embedded in its core.

## Optical Characterization

The optical transport through the fibre and the coupling efficiencies to the embedded SiPMs were measured using a 405 nm, nanosecond-pulsed laser source that perpendicularly excites the scintillator fibre a certain distance away from the embedded SiPM **(Figure 2a)**. A fraction of the incident laser power was split to an external photodetector to correlate the timing of the current pulses received on the embedded SiPM to the incident laser pulses, allowing us to collect only those current pulses that are coincident with the laser source. The electrical pulses measured from the SiPM show distinguishable quantized levels corresponding to the number of microcells that undergo avalanche due to simultaneous photon detection events on separate pixels of the SiPM **(Figure 2b)**.[33,34] As such, a histogram of the integrated pulse areas, which are proportional to the charge collected per event, over many laser pulse events reveals distinct peaks corresponding to discrete numbers of photoelectrons detected in the SiPM **(Figure S6)**. The average number of detected photoelectrons is directly related to the number of incident photons through the photon detection efficiency of the device. Despite electromagnetic interference and parasitic impedance introduced by the active fibre electrodes, the width of the peaks in the spectrum remains narrow relative to their separation, enabling resolution of individual photon detection events in the fibre. The position of the maxima of these peaks shows a high degree of linearity with the peak number **(Inset Figure 2b)**, allowing us to calibrate the pulse integration area to the number of photoelectrons detected in the fibre.

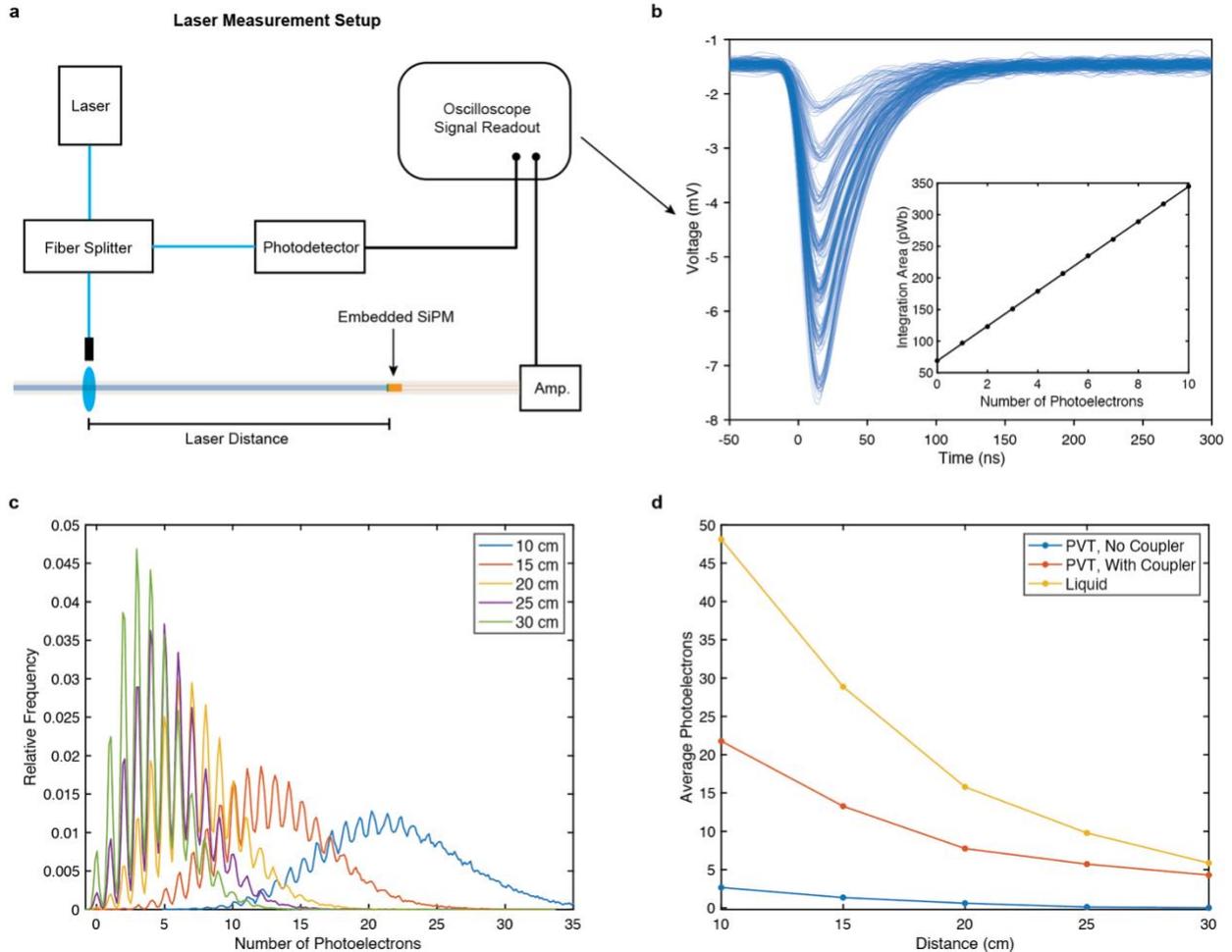

**Figure 2: Optical Characterization | a.** Schematic of the laser setup used to characterize the optical transport through the different types of scintillating fibres. A pulsed laser source is incident on a localized section of the fibre a certain distance away from the embedded SiPM while the corresponding pulse waveforms are measured on the oscilloscope. A portion of the incident laser signal is split to an external photodetector to collect only those pulses on the SiPM that correspond to an optical pulse from the incident laser source. **b.** Overlay of many pulse waveforms received on the SiPM in response to the incident laser source, showing distinct levels corresponding to discrete numbers of photoelectrons detected in each pulse. A histogram of the integrated areas of each pulse shows distinct photoelectron peaks **(Figure S6)**, the positions of which have a linear relationship with the number of photoelectrons detected in each pulse (inset). **c.** The photoelectron spectra measured on the plastic scintillator fibre for different distances of the pulsed laser source from the embedded SiPM, with distinct photoelectron peaks in each spectrum. **d.** The average number of photoelectrons detected per laser pulse for different distances between the laser source and embedded SiPM. The attenuation characteristics are compared for a plastic scintillator fibre without an optical coupler, plastic scintillator fibre with an optical coupler, and a liquid scintillator fibre. The

comparison shows that the coupler improves light collection by a factor of ~10x, with transmission in the liquid scintillator fibre being the highest.

At each laser distance, the measured photoelectron spectrum follows a Poisson-like distribution due to the discrete probabilistic nature of photon emission, transport, and detection, with the finite-width photoelectron peaks imposed on the distribution **(Figure 2c, Figure S7)**. As the laser is moved along the length of the fibre from 10 cm to 30 cm away from the embedded SiPM, the distribution shifts to lower numbers of photoelectrons detected per laser pulse **(Figure 2d)**, due to attenuation of light from scattering, absorption, or potential leakage of quasi-guided modes. Before the injection of the optical coupler, the plastic scintillator fibre shows relatively poor coupling between the PVT waveguide and SiPM, given that the SiPM interfaces axially with the hollow air channel in which the guided modes of the waveguide are evanescent. The introduction of the optical coupler in the plastic scintillator fibre leads to a nearly 10-fold improvement in the average number of photoelectrons detected per laser pulse, showing the efficacy of the coupler in enhancing the coupling efficiency from the annular waveguide to the SiPM. Compared to the plastic scintillator fibre, the liquid scintillator fibre showed greater scintillation yield with respect to the incident 405 nm laser source or a higher coupling efficiency, but also greater attenuation losses with distance. By fitting an exponential to the attenuation curve of the average photoelectrons received per pulse, we find that the plastic scintillator fibre shows an attenuation length of $12 \pm 1$ cm compared to $9.4 \pm 0.2$ cm for the liquid scintillator fibre. This is likely due to the lower index contrast between the liquid scintillator and ECOC cladding, which can lead to greater losses in the more weakly confined guided or quasi-guided modes. In the future, improvements to the smoothness of optical interfaces, material transparency, and index-contrast of the core-cladding materials could help lead to lower attenuation losses. For example, further refinements may be needed to minimize any irregularities that could form on the inner surfaces of the fibre when drawing metallic electrodes within a confined polymer melt, as these may contribute to additional scattering losses. Additionally, alternate light-guiding mechanisms, such as photonic bandgap structures,[40] could be introduced to significantly enhance the fraction of isotropically-emitted photons that are captured and guided through the scintillator.

## Radiation Detection

After characterizing the ability of the fibre to transport, couple, and detect scintillation light produced along its length, we measured the response of the fibre to three different radiation sources: a 0.5 μCi Sr-90 beta source, and two 10 μCi gamma sources of different energies, Cs-137 and Co-60. These sources produce 0-1.1 MeV electrons, 0.662 MeV gammas, and a pair of 1.17 and 1.33 MeV gammas, respectively. The beta source is primarily used here to first provide a direct charged-particle interaction, enabling fast, efficient, and statistically robust characterization of the detector response, while also serving as a practical proxy for the effect of secondary electrons produced by gamma interactions. It should be noted that given the ~mm thickness of the scintillator, only a small fraction of gammas interact with the fibre via Compton scattering, and the resulting Compton electrons typically deposit less than half of their energy in the scintillator volume.

To measure the response of the fibres to these three sources, we collimated the emission of sealed disk source standards using plastic (for electrons) or lead (for gammas) shielding with narrow openings to perpendicularly expose a highly localized section of the fibre a certain axial distance away from the embedded SiPM **(Figure 3a)**. This configuration isolates the contribution of scintillation events generated at specific positions along the fibre and allows us to probe the axial transport of scintillation light to the detector. Because the inverse-square dependence on radial distance is well-characterized, we keep the radial source distance constant in all measurements. Unlike the laser measurements, which were performed in coincidence between the laser and the SiPM signal, here, we trigger directly on the electrical pulse produced by the SiPM, as would be the case in a realistic application. Because of this, we must now also deal with the interference of thermally-generated charge carriers, or dark counts, in the SiPM that do not correlate to photon detection events but produce charge pulses indistinguishable from photo-generated carriers.[33,34] Note that for ease of characterization, radiation measurements were performed in a dark box to avoid ambient light from also generating interfering signals within the SiPM. However, we find that thin, black polyolefin tubing conforming to the profile of the fibre can provide the same level of ambient light suppression as the dark box **(Figure S8)**. Given that carbon-black-loaded polymers have already been shown to be compatible with the thermal draw process,[28,38] in the future, these highly light-absorbent materials could be co-drawn as a light-tight cladding for the fibres we show in this work.

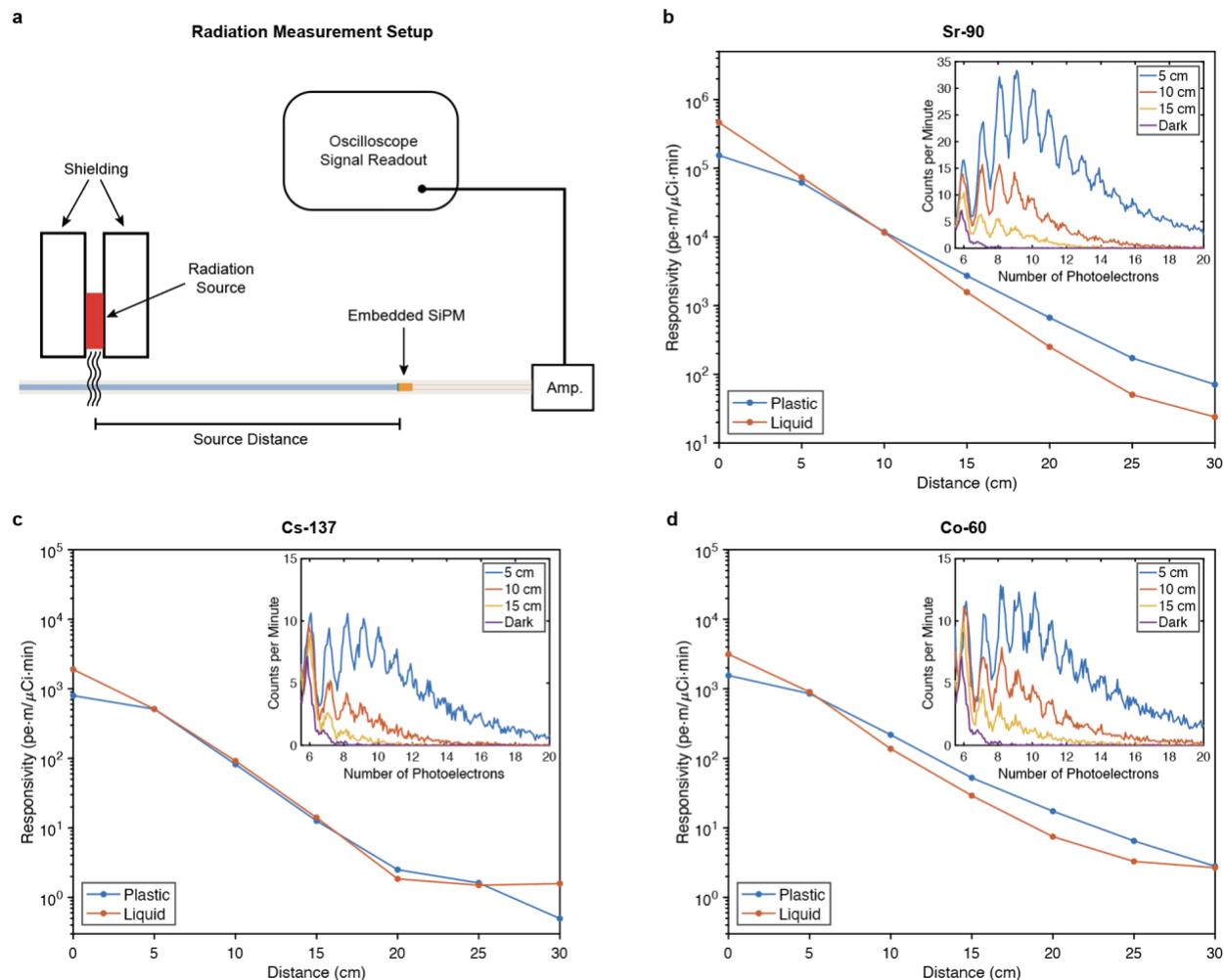

**Figure 3: Radiation Detection | a.** Schematic of the setup used to characterize the response of the plastic and liquid scintillating fibres to different radiation sources. The emission of the source standards were shielded and collimated to locally expose the fibre a certain distance away from the embedded SiPM while pulse waveforms are measured on an oscilloscope. Responsivity of the plastic and liquid scintillator fibres to a **b.** Sr-90 beta source, **c.** Cs-137 gamma source, and **d.** Co-60 gamma source at different distances along its length from the embedded SiPM. Responsivity curves were normalized based on experimental responses to the collimated radiation sources **(Figure S11)**. The inset of each plot shows examples of the photoelectron spectrum measured for the plastic scintillator fibre when the collimated sources were 5 cm, 10 cm, and 15 cm away from the embedded SiPM, with the dark photoelectron spectrum indicating the count rate measured in the absence of a radiation source.

As with the laser, distinct photoelectron peaks can be seen in the fibre's response to the collimated radiation sources **(Inset Figure 3b-d, Figure S10)**. However, thermally-generated electron-hole pair events introduce an additional dark count distribution that contributes to the total measured response. For low photoelectron peaks, the dark count rate dominates, with little contrast between count rates measured with and without radiation **(Figure S9)**. Because the probability of simultaneously triggering multiple adjacent microcells via optical crosstalk decreases exponentially with each additional cell, the dark count rate is increasingly suppressed for higher photoelectron peaks, enabling clearer discrimination of radiation-induced events. To improve the signal-to-noise ratio, we applied a photoelectron threshold of 8 photoelectrons to filter out the majority of the dark count contributions. As the distance between the source and embedded SiPM increases, the radiation response distribution shifts towards lower numbers of photoelectrons detected, becoming increasingly dominated by the dark count response below the photoelectron threshold. Along with the optical losses characterized earlier, this interference from the dark counts and the need for a filtering threshold provides an additional contribution to the attenuation in the total photoelectron detection rate. Using the measurement setup, we observe a significant response to the collimated radiation sources above the dark count level at distances of up to ~30 cm **(Figure S11)**.

To compare the responses to different sources, we normalized the measured photoelectron detection rates by the estimated collimated source intensity to generate responsivity curves that indicate the sensitivity of each point along the fibre to local field intensity **(Figure 3b-d)**. The responsivity of the fibres is highest for the Sr-90 beta source, given that it involves direct interaction with ~MeV electrons. For gamma sources, such as Cs-137 or Co-60, a secondary scattering process, such as Compton scattering, is required to produce sub-MeV electrons in the scintillator, reducing the probability of interaction with the fibre.[41] Between the gamma sources, the Co-60 source produces a larger optical response than the Cs-137 source, given that the higher-energy Co-60 gamma rays deposit more energy and thus yield a higher number of scintillation photons per interaction event with the fibre.

Comparing the plastic and liquid scintillator fibres, we find results consistent with the laser measurements discussed earlier. For each source, close to the embedded SiPM, the liquid scintillator fibre exhibits a higher responsivity due to its higher scintillation yield and more favourable coupling geometry. Namely, the cylindrical core-cladding structure of the liquid scintillator fibre allows the scintillating material to directly contact the face of the SiPM, enabling a large angular acceptance range for photons produced near the embedded SiPM. Meanwhile, the plastic PVT scintillator fibre uses an annular scintillating domain that does not directly contact the active face of the SiPM, requiring photons to travel through a separate optical coupling region that limits the angular acceptance range and introduces losses from interfacial scattering, reflection, or absorption. As the source is moved further away, the effect of optical transport losses begins to dominate over coupling losses. In this regime, we find that the plastic scintillator fibre generally produces a stronger signal than the liquid scintillator fibre due to its longer light attenuation length (**Figure 2d**).

The experimental responsivity measurements can be integrated to model the response of the fibres to arbitrary radiation fields, such as uniform fields (**Figure S12a**) or spatially varying fields from localized emitters (**Figure S12b**). In the case of a uniform field, this analysis allows us to estimate practical lower gamma detection limits that produce responses distinguishable from the experimentally measured variation in the dark count noise level (**Figure S12a**). Depending on the source and fibre type, we find that these estimated detection limits range from ~14-41 nSv/hr (**Table S1**), with the liquid scintillator fibre having slightly lower detection limits due to its more favourable coupling geometry. The fibres' response to localized emitters reveals the benefits of the fibre's distributed interaction area (**Figure S12b**). Because optical waveguiding transports scintillation light generated along the fibre to the embedded SiPM, the fibre acts as an extended detector that integrates contributions over its length. As a result, the total response of the fibre to localized emitters exhibits a more gradual falloff with axial distance than the inverse-square dependence that would be observed with a point detector **(Figure S12b)**.

The localization of transport and detection within each fibre offers a straightforward solution for enabling radiation sensitivity across arbitrarily long lengths. By embedding multiple SiPMs at spacings less than the maximum detectable range of a single SiPM, the architecture can remain fully responsive throughout. This integrated optoelectronic approach ultimately surpasses the attenuation limits of externally coupled systems, providing a scalable pathway to wide-area radiation detection.

## Mechanical Characterization and Fabric Formation

To demonstrate the viability of a large-area, conformal fabric detector, we show that the fibres possess the requisite mechanical properties needed to robustly withstand the rigors of fabric formation and usage. Here, we will focus on the unique advantages of the liquid scintillator fibre, which uses a low-modulus elastomeric cladding that helps to create a substantially more flexible and stretchable fibre compared to those composed of conventional thermoplastics. In addition to the intrinsic material elasticity of the ECOC, which can stretch over 450% without breakage, structural elasticity can be built into the electrically-active side of the fibre by inducing a buckling instability of the copper wires post-draw, or by plastically deforming the copper wires into a helical structure pre-draw.[39,42] This approach has been shown to allow the fibre to tolerate elastic deformations greater than 60% while maintaining electronic functionality. In **Figure 4a-b**, we also show that stretching of this elastic liquid scintillator fibre has minimal impact on the optical detection capabilities, with 50% stretching producing nearly the same photoelectron spectrum as the unstretched fibre upon laser excitation.

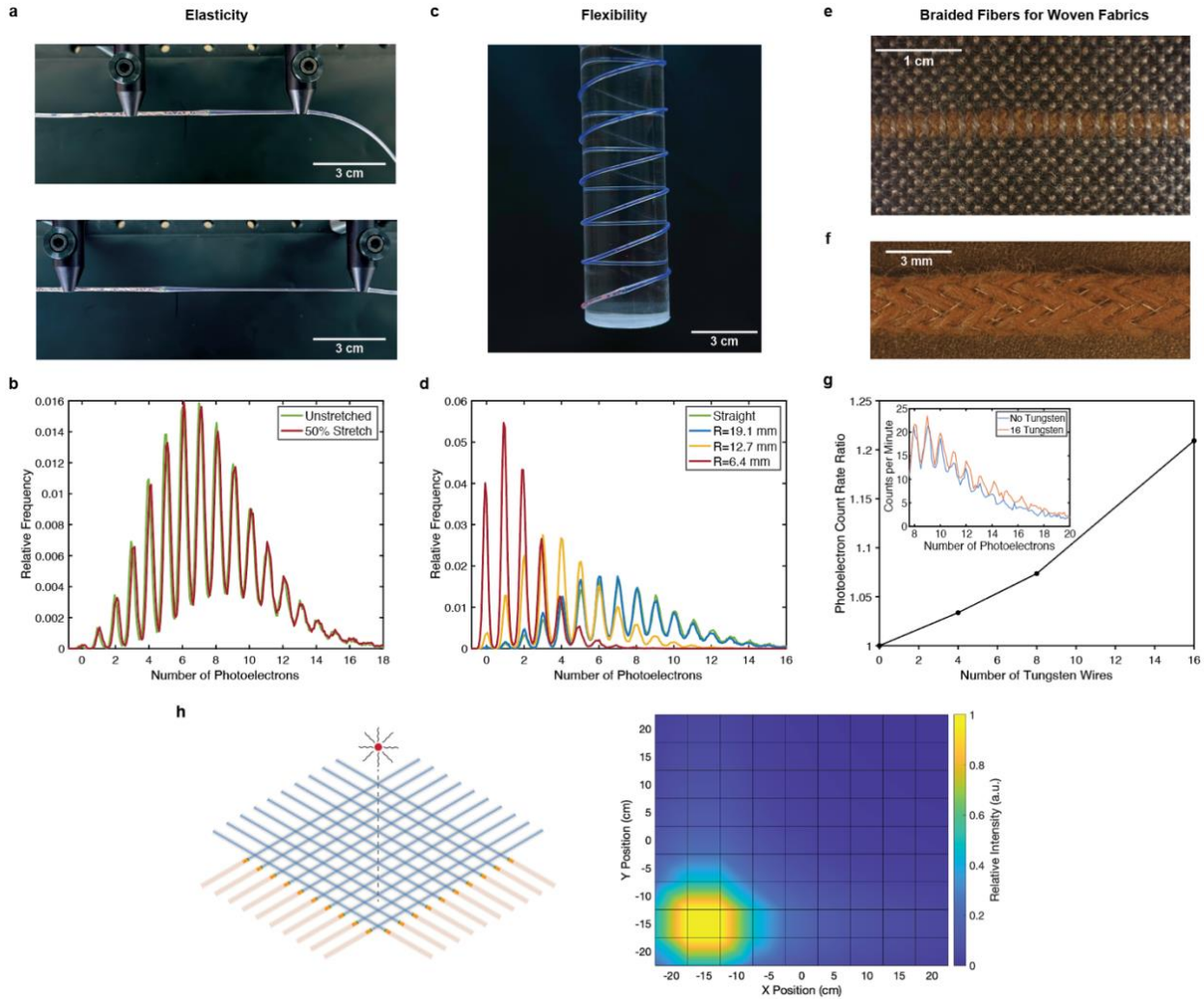

**Figure 4: Mechanical Characteristics and Fabric Formation** | **a.** Images of the liquid scintillator fibre in its unstretched state (top) and after being stretched by over 50% (bottom). **b.** The photoelectron spectra measured on the liquid scintillator fibre upon pulsed laser exposure in its unstretched state and while being stretched by 50%, showing minimal effect of stretching on the optical performance of the fibre. **c.** Image of the liquid scintillator fibre being wrapped around a cylindrical mandrel with a radius of curvature of 19.1 mm. **d.** The photoelectron spectra measured on the liquid scintillator fibre upon pulsed laser exposure while straight, and while wrapped around mandrels with various radius of curvatures. Optical losses are minimal down to bend radii of 19.1 mm, below which we begin to see shifting of the photoelectron spectra to lower numbers of photoelectrons collected. **e.** Image of a liquid scintillator fibre that was braided with merino wool and machine woven into a fabric via a standard loom. The portion of the fabric shown in the photograph includes the section of the fibre with the embedded SiPM. **f.** Close-up image of a fibre braided with a combination of merino wool and 100 μm tungsten wires. **g.** Effect of the number of 100 μm tungsten wires braided around a plastic scintillator fibre on the total photoelectron count rate measured upon exposure to a collimated Co-60 source, normalized to that of the fibre braided only with merino wool. The inset compares the photoelectron spectra measured for the fibre braided with 0 and 16 tungsten wires. **h.** Simulation of the response of a 10 x 10 woven array of plastic scintillator fibres to a Co-60 point source above the fabric, showing the potential for such woven fabrics to map spatially-varying radiation fields over large areas.

We also showed the ability of the fibre to withstand the bending deformations needed to form fabrics. Electrically, these fibres have been shown to be capable of bending radii down to ~2-3 mm near the embedded device, and 0.25 mm away from the embedded device.[39] However, because tight bend radii can lead to leakage of the optical modes traveling through the fibre, the impact of the fibre curvature on the optical performance also places a constraint on the minimum bend radii that can be sustained.[43] In **Figure 4c-d**, we show how the intensity received at the embedded SiPM attenuates as a function of the bend radius the fibre is wrapped around. For a radius of curvature of 19.1 mm, the photoelectron spectrum of the wrapped fibre nearly matches that of the fibre when straight, indicating that bending losses are minimal even down to a bending radii of 19.1 mm. However, below this bend radius, we begin to observe shifting of the photoelectron spectrum to lower numbers of photoelectrons due to optical bending losses.

Given the unique flexibility and stretchability of these fibres, we show that they can be used as a viable stock yarn for direct assembly into fabrics. To do so, we first cover the outer surface of the fibres with a textile braid consisting of diagonally interlaced yarns. The braid acts to both amplify the tensile strength of the core fibre, and also provide a low-friction textile surface that minimizes stiction during fabric formation.[44,45] In **Figure 4e**, we show an example of a liquid scintillator fibre braided with merino wool that was directly woven into a fabric using a standard loom. The braid helps to match the mechanical and surface properties of the fibre to the carrier yarn material of the fabric, allowing it to weave naturally without causing distortion in the fabric. Beyond fabric construction and durability improvements, the braid also grants an additional degree of freedom towards improving the fibre's sensitivity to radiation. High-Z materials, such as tungsten, interact more strongly with gamma radiation than low-Z materials, such as organic scintillators.[46] Therefore, adding heavy metals around the fibre can increase the generation rate of secondary electrons that deposit energy in the fibre's scintillating material.[47,48] The braid provides a natural platform to do so, given its capability to incorporate metallic wires into its structure. In **Figure 4f**, we show a plastic scintillator fibre braided with a combination of merino wool and 100 μm tungsten wires. As expected, we find that increasing the number of tungsten wires incorporated in the braid enhances the photoelectron detection rate upon Co-60 exposure, with an over 20% increase observed for a braid containing 16 tungsten wires (**Figure 4g**). To yield even more drastic improvements in the future, the wire diameter, braid density, and braid composition can be tuned to further enhance gamma interaction.

To illustrate the potential utility of fabrics constructed of fibre detectors, we simulated the response of a 50 x 50 cm woven fabric with 10 plastic scintillator fibres in both the warp and weft direction to a Co-60 point source, as shown in **Figure 4h**. Using the experimentally measured responsivity of the individual fibre detectors, we modelled the signals that would be detected by each fibre and combined them at each fibre crossing to reconstruct an intensity map of the detected radiation field across the fabric. This simulation shows that the distributed woven array of fibres can capture the emission pattern of the point source across a large-area fabric. More broadly, this example illustrates how single-fibre detectors could serve as the fundamental unit for conformal, large-area distributed sensing architectures capable of real-time, high spatial-resolution radiation mapping with high sensitivity, pointing toward the future realization of fabric-scale radiation detection.

## Conclusion

In summary, we demonstrate flexible and elastic optoelectronic fibres capable of real-time, distributed radiation detection of beta and gamma radiation in the 0.1-1.5 MeV range with single-photon resolution. By pursuing various architectures and compositions, we show that the method is quite general and can be tailored to meet particular optical and mechanical requirements. Braiding the fibre allows one to increase the phase space of possibilities to enable high Z detection enhancement and textile covering for fabric incorporation. The estimated detection limits of these fibres (~14-41 nSv/hr) approach near-background radiation levels, highlighting their potential applicability for gamma dosimetry in low-to-moderate radiation environments, such as on-body monitoring in human-occupied spaces.

While the present work establishes proof-of-concept performance, the long-term stability and radiation tolerance of both the embedded SiPMs and scintillators will ultimately define the upper useable dose-rate limit. The performance of organic scintillators and SiPMs are known to degrade under high radiation fluence, as well as environmental stressors such as temperature and humidity cycling, which may affect light output, dark count noise, and sensitivity [REF]. These effects scale with cumulative dose and dose rate and become most significant in harsh radiation environments. Therefore, future application-specific testing under such conditions will be necessary to validate long-term performance, especially for higher-dose or safety-critical deployments.

Building on our work, numerous advancements can be made at the single fibre level with time. While the current fibre has a millimetre-scale diameter, it is limited by the size of the SiPM. Smaller arrays of single-photon avalanche photodiodes can be used to facilitate further fibre miniaturization.[49] Additionally, incorporation of in-fibre computing and wireless communication capabilities alongside the established detection capabilities can help create a completely untethered system for real-time readout without any external electronic circuitry.[39] In addition, future mobile implementations may benefit from integrated temperature sensing within the fibre [REF] combined with passive calibration or active bias compensation to correct for shifts in SiPM gain due to environmental fluctuations [REF].

At the fabric level, extrapolation of experimental single-fibre responsivity measurements indicate that multi-fibre arrays could provide a route toward highly conformal, spatially-resolved textile detectors or other distributed sensing architectures. Beyond field mapping, advanced algorithms and networking capabilities could enable coincidence measurements between multiple fibres, allowing for source localization, directionality, particle identification, energy reconstruction, and recoil tracking.[50–53] Future advancements in scalable, synchronized multi-fibre readout and detection in large-area textiles will be key to realizing such fabric systems.

Beyond gamma dosimetry for personal or environmental monitoring, this type of distributed, real-time fabric detection system could be advanced for a variety of other applications in the future, such as high-energy physics or nuclear medicine.[54–59] For example, the integrated optoelectronic fibre architecture presented here, with photodetectors embedded directly within the fibre, could simplify the construction of large, robust detector arrays for high-energy physics compared to conventional systems in which scintillating fibres are externally coupled to photodetectors. In such contexts, additional design factors may become important, such as particle discrimination or radiation hardness. To this end, future implementations could incorporate dual-readout schemes, for example combining fibres that are composed of materials sensitive to different types of radiation, or employing particle-specific shielding to enable particle discrimination. Ultimately, continued advancements at both the single-fibre and fabric levels will expand the versatility of these fibres, enabling a wide range of applications in both gamma dosimetry and future distributed radiation detection platforms.

# Methods

**Radial-to-Axial Device Conversion**
To minimize the fibre dimensions, we chose to integrate a 1600-microcell SiPM (AFBR-S4K11C0125B; Broadcom) with a 1 x 1 mm$^2$ active area, SPAD pitch of 25 μm, and overall dimensions of 1.315 x 1.315 x 0.595 mm$^3$. First, the ends of two enamelled copper wires (38 AWG; Remington) were soldered to the exposed anode and cathode contact pads on the back face of the device, with the polyurethane insulation preventing shorting between the conductors. The wires were then plastically deformed to extend perpendicularly from the back face of the device. To hold tension on the side of the device with the active photodetector array, an enamelled copper wire (38 AWG; Remington) was looped around the back-side of the device without being electrically connected to any of the contact pads. The device was then pulled tightly into the hole of a Teflon mould, whose opening was nearly matched to the planar dimensions of the chip (1.3 x 1.3 mm), but whose depth was designed to add 2 mm to the device's thickness, which was previously 0.6 mm. The empty space of the mould was then filled with a liquid adhesive (NOA 61; Norland), which was UV-cured to the wires and back-face of the device. After curing, the device was pushed out of the mould to reveal the elongated SiPM, now 2.6 mm in thickness. Many devices were moulded in this way, and their electrodes were connected together a few metres apart on a single linear chain to enable drawing of multiple SiPMs through a single preform.

**Rheological Measurements**
Oscillatory shear viscosity measurements were carried out using a rotational rheometer (Discovery HR 20; TA Instruments). Polymer samples were placed between two 25 mm diameter parallel plates, with a plate separation of 2 mm. The samples were heated up to 250°C in an environmental testing chamber, and the rheological properties measured while the temperature was decreased from 250°C to 100°C at 3° C/min. The rheological characteristics were measured in oscillatory shear mode at a frequency of 1 rad/s and a strain amplitude of 1%.

**Plastic Scintillator Fibre Draw Process**
The plastic scintillator fibre was drawn from a macroscopic rectangular preform consisting of a 3/4 in. x 3/4 in. PVT scintillator (EJ-200; Eljen) encased by a 1 in. x 1 in. PMMA cladding, with a 3/8 in. x 3/8 in. hollow channel in the centre of the preform. To fabricate the preform, CNC milling was used to cut a 3/8 in. x 3/16 in. rectangular trench in a 3/4 in. x 3/8 in. bar of PVT. Similarly, a 3/4 in. x 3/8 in. rectangular trench was milled in a 1 in. x 1/2 in. bar of PMMA. The PVT bar was then nested inside the milled PMMA bar to form one half of the preform. Two halves made in this way were then consolidated together in a hot press at 5 psi and 120 °C for 1 hour, with a removable 3/8 in. x 3/8 in. Teflon bar filling the air channel to prevent collapse during consolidation. During the draw, the preform was hung in a three-zone furnace set to an upper temperature of 100 °C, middle temperature of 265 °C, and lower temperature of 75 °C. Under the tension of the pulling capstan, the preform necked and flowed to fibre dimensions while the pre-connected SiPMs were fed into the single large channel of the preform. To enable a draw down ratio of approximately 12, a preform down feed speed of 1 mm/min and a capstan speed of 0.15 m/min was used, while monitoring the resulting diameter using a laser micrometre. Once the devices were drawn into the core of the fibre, an optical coupler was introduced directly adjacent to the devices post-draw. To do this, stainless steel tubing with an outer diameter of 0.018 in. was inserted into the air channel of the fibre such that one end of the tubing was directly adjacent to the photodetector face of the SiPM, while the other end of the tubing was connected to a 0.25 mL syringe filled with a liquid photopolymer (NOA 61; Norland). A microfluidic syringe pump (Genie Plus; Kent Scientific) was used to inject the liquid photopolymer at a rate of 5 μL/min, until the liquid filled a 5 mm length of the air channel directly adjacent to the device. The tubing was removed from the fibre, and the liquid photopolymer was UV-cured until it formed a solid coupling region near the device.

**Liquid Scintillator Fibre Draw Process**
The pre-connected SiPMs were thermally drawn into the core of an ECOC fibre by adapting the same draw process described in previous work.[39] After the draw process, the fibre consists of a cylindrical air core surrounded by a cylindrical ECOC cladding, with the active photodetector array of the embedded SiPM facing the air core of the fibre. On the optically active side of the fibre, the mechanical guide wires were removed by pinching the deformable ECOC cladding to pin the wires at the SiPM interface, and then pulling the ends of the wires from the fibre end to break them near the device and extract them from the fibre. Then, the optically active side of the fibre was filled with a liquid scintillator (EJ-309; Eljen). To do so, 0.018 in. diameter stainless steel tubing was inserted into the air channel of the fibre until its tip was directly adjacent to the device. The other end of the tubing was coupled to a 0.25 mL glass syringe, which was filled with the liquid scintillator. The liquid was injected into the fibre at a rate of 15 μL/min using a microfluidic syringe pump (Genie Plus; Kent Scientific) until the liquid scintillator fully filled the air channel from the device to the end of the fibre. The end of the fibre was then sealed using a UV-cure adhesive (NOA 61; Norland) to encapsulate the liquid scintillator within the fibre channel.

**Fibre Signal Readout**
The in-fibre electrodes were exposed at the electrically-active end of the fibre by melting off a short length of the polymer cladding, and then soldered to a two-pin header connector. To readout signals from the fibre, the end of the fibre was connected to an amplifier PCB (AFBR-S4E001; Broadcom) that connects the embedded fibre SiPM to a circuit consisting of a high-voltage filter, transimpedance amplifier, and buffer with unity gain. The amplifier board was powered using a +/- 5 V dual power supply (DP831; Rigol), while the SiPM was biased at 27.5 V using a high-voltage power supply (E3612A; Agilent). The output of the amplifier PCB was connected to an oscilloscope (HDO6104; Teledyne Lecroy) operating at 200 MHz bandwidth, which acts to integrate and histogram the SiPM pulse waveforms in real-time using a 50 ns integration window and 8 ns offset from the pulse trigger. All laser and radiation measurements performed on the fibre were done in a black, wooden dark box to suppress ambient light levels. During measurements, the fibre and amplifier board were suspended horizontally inside the dark box, and connected to the external power supplies and oscilloscope via electrical feedthrough cables. All measurements were performed under controlled laboratory conditions, with ambient temperature maintained at approximately 22°C.

**Optical Measurements**
The optical transport characteristics through the fibres were measured using a 405 nm pulsed laser source (NPL41C) with a nominal pulse width of 6 ns. The laser was triggered externally at 50 kHz using a function generator (SDG 1032X; Sigilent). The output of the laser was coupled into a fibre optic cable using a FibrePort coupler (PAF-X-18-PC-A; ThorLabs), with an adjustable mirror setup used to tune the laser coupling into the fibre. The laser signal was then split into two fibre optic cables using a 1x2 fibre optic coupler (TM105R5F1A; ThorLabs). One of the fibre optic cables was fed into a silicon avalanche photodetector module (APD130A2; ThorLabs) that was electrically connected to the oscilloscope and used as a trigger to correlate the electrical pulses received from the embedded SiPM fibre to the incident laser source. The other fibre optic cable was mounted on a movable rail, such that the end of the fibre optic cable was facing perpendicularly to the scintillating fibre with the embedded SiPM. This fibre-coupled pulsed laser was used to excite the scintillating fibre at different distances away from the embedded SiPM while the area of the electrical pulses received from the fibre were histogrammed in real-time on the oscilloscope.

**Bend Testing**
The laser measurement setup described above was used to determine the influence of bending on optical losses through the liquid scintillator fibre. To do so, the fibre-coupled pulsed laser was positioned a certain distance away from the embedded SiPM in the scintillator fibre. The response of the scintillator fibre to the pulsed laser source was first measured in this straight configuration. Then, while holding the position of the fibre-coupled laser constant, the entire length of scintillator fibre preceding the laser incidence region was wrapped around a cylindrical mandrel, and the response of the scintillator fibre to the laser source in this bent configuration was measured. This was repeated for cylindrical mandrels of different diameters to determine the influence of the bending radius on the optical signal received at the embedded SiPM.

**Tensile Testing**
The laser measurement setup was also used to determine the influence of stretching on the optical response of the liquid scintillator fibre. For these measurements, the liquid scintillator fibre was clamped at the location of the embedded SiPM and at the optically-active end of the fibre, such that the optically-active length of the fibre was suspended horizontally. The fibre-coupled laser was positioned between these two clamps a certain distance away from the embedded SiPM in the scintillator fibre. The response of the fibre to the laser source in the unstretched configuration was first measured. Then, the fibre was stretched to 50% tensile strain by horizontally pulling the clamped end of the fibre while keeping the position of the SiPM clamp constant. The response of the scintillator fibre to the pulsed laser source was measured in this stretched configuration. While stretched, the clamps were firmly mounted in optical breadboards to prevent the tension of the stretched fibre from relaxing the tensile strain.

**Radiation Measurements**
We measured the response of the fibres to three different radiation sources: a 0.53 µCi Sr-90 beta source, and two 10.4 µCi gamma sources, Cs-137 and Co-60. To localize exposure of the source to certain sections of the fibres, each source was collimated in a way that was appropriate for the geometry of the sealed source standard. The Sr-90 source standard (Isotope Products Laboratories) was distributed uniformly over the surface of a disk, with an active diameter of 20.4 mm. To collimate its emission, we encapsulated the source in a 6.35 mm thick acrylic shield and drilled a 6.35 mm diameter hole in the centre to allow for localized exposure of the fibre directly above the hole. Meanwhile, the Cs-137 and Co-60 source standards (Isotope Products Laboratories) each contained a 5 mm diameter cylindrical active element sealed in a plastic disk. To collimate their gamma emission, they were placed between two 5 cm thick lead blocks to localize the exposure of the fibre to only the opening of the blocks, which was approximately 6 mm. During measurements, the distance from the centre of the active element in the source to the fibre was approximately 15 mm. The response of the fibres to perpendicular exposure from the collimated sources was measured for different distances away from the embedded SiPM, using the oscilloscope for pulse waveform collection and analysis. At each distance, pulse waveforms were collected for 30 minutes to generate sufficient resolution in the photoelectron spectrum. Unlike the laser measurements, the oscilloscope here was triggered directly on the signal received from the embedded SiPM, so the measured response also includes the influence of spontaneous thermally-generated photoelectrons that contribute to dark counts even in the absence of radiation. To minimize the interference of the dark counts, the total photoelectron count rate, $f$, above a photoelectron threshold of 7.6 photoelectrons was calculated for each photoelectron spectrum **(Figure 3b-d, Figure S10)** using a weighted sum,

$$f = \sum_{p=7.6}^{\infty} p * N(p) \quad (1)$$

where $p$ is the number of photoelectrons and $N(p)$ is the count rate for a certain number of photoelectrons in the spectrum. The results for the total photoelectron count rates at each distance are plotted in **Figure**

**S11**. To estimate the variance of the dark count rate in our setup, we measured the photoelectron spectrum in the absence of a radiation source 8 separate times. The mean and standard deviation of the dark photoelectron count rate across these 8 measurements were used to estimate a prediction interval using,

$$PI = f_{dark} \pm t_{0.05,7} * s_{dark} \cdot \sqrt{1 + \frac{1}{n}} \qquad (2)$$

where $f_{dark}$ is the mean dark photoelectron count rate (61 photoelectrons per minute), $s_{dark}$ is the standard deviation of the dark photoelectron count rate (11 photoelectrons per minute), $t_{0.95,7}$ is the one-tailed critical t-value for 7 degrees of freedom at the 5% significance level, and $n$ is the number of dark count measurements (8). This interval is represented by the red area in **Figure S11**, above which the count rate differs from the dark count rate at the 5% significance level.

**Responsivity Curves**
Normalizing the response of the fibres to the intensity of collimated radiation sources can be used to generate responsivity curves that indicate the sensitivity of the fibres to local field intensity along their length. Given the geometry of the collimation setup, we approximated the Co-60 and Cs-137 source standards as point source emitters with nearly uniform intensity over the localized section of fibre exposed. We also assume that the lead shielding fully attenuates all other gamma emission except that which is collimated between the opening in the shielding. With these assumptions, the responsivity of the fibre at different positions along its length, $R(z)$, could be found by normalizing the experimental measurements to the incident source intensity and length of fibre exposed,

$$R(z) = \frac{f - f_{dark}}{\left(\frac{A}{4\pi d^2}\right) l} \qquad (3)$$

where $f$ is the measured photoelectron count rate, $f_{dark}$ is the mean dark photoelectron count rate in the absence of a radiation source, $A$ is the activity of the source used in the experimental setup (10.4 µCi), $d$ is the distance from the center of the source to the fibre (15 mm), $l$ is the length of the fibre exposed through the opening in the lead shielding (6 mm), and $z$ is the axial distance from the embedded SiPM. The estimated incident source intensity at the fibre for the two gamma emitters was approximately 0.4 µCi/cm². Using the gamma ray dose constants for Cs-137 and Co-60 (0.33 R/hr and 1.32 R/hr, respectively, for a 1 Ci source at 1 m), this corresponds to dose rates of 153 µSv/hr and 610 µSv/hr, respectively. Meanwhile, the Sr-90 source standard was modelled as a uniform surface-emitting disk source whose active area is limited by the hole in the center of the acrylic shielding. Similar to before, we assume that the intensity of the source is nearly uniform over the section of fibre that is exposed, and that the plastic shielding fully attenuates all other beta emission besides that collimated through the hole in the shielding. With these assumptions, we calculate the normalized responsivity of the fibre to the Sr-90 source using,

$$R(z) = \frac{f - f_{dark}}{\left(\frac{A}{4\pi R^2} \ln\left(\frac{d^2 + r_c^2}{d^2}\right)\right) l} \qquad (4)$$

where $A$ is the total activity of the Sr-90 source used in the experimental setup (0.53 µCi), $R$ is the total radius of the disk source (10.2 mm), $d$ is the distance between the source and the fibre (9.5 mm), $r_c$ is the active source radius limited by the hole in the shielding (4.8 mm), and $l$ is the length of the fibre exposed (6.4 mm). The resulting responsivity curves for the liquid scintillator fibre and plastic scintillator fibre are shown in **Figure 3b-d**.

## Fabric Response Simulation

Woven arrays of fibres can be used to map spatially-varying radiation fields. In **Figure 4h**, we simulate the response of a 10 x 10 woven array of plastic scintillator fibres separated by 5 cm each to a 10 µCi Co-60 point source positioned 5 cm above the fabric. The signal detected by each fibre was modelled by integrating the plastic scintillator fibre's responsivity curve to Co-60, shown in **Figure 3d**, according to Equation S3. We then computed the outer product of the weft and warp fibre response arrays, $r_{weft}$ and $r_{warp}$, to find the matrix of signal intensities, $S$, at each fibre crossing in the woven array,

$$S = r_{weft} \otimes r_{warp} \tag{5}$$

By linearly interpolating between the fibre crossings, we generated a reconstructed map of the detected signal intensity across the fabric.

## Braiding

Fibres were braided using a 16-carrier braiding machine (16B-E80; Ratera). Each carrier was loaded with either superfine merino wool yarn (2/18; JaggerSpun), or a 100 µm tungsten wire (CAS 7440-33-7; Alfa Aesar) twisted in with the merino wool yarn. The amount of tungsten incorporated in the braids was controlled by adjusting the number of carriers that had tungsten wire twisted in with the merino wool yarn. During braiding, the horn gear speed and take up speed were adjusted to keep the braid density within 6-8 stitches per cm.

## Machine Weaving

Fibres were machine-woven into fabrics using an AVL Compu-Dobby loom. Each warp line consisted of two cotton yarns at a density of about 25 ends per inch. The braided scintillator fibres were woven into the weft direction of the fabric, which otherwise consisted of a 2/8 wool yarn woven at 20-25 yarns per inch.

# References

1. Kouzes, R. Radiation Detection Technology for Homeland Security. in *Handbook of Particle Detection and Imaging* 897–927 (Springer International Publishing, 2020). doi:10.1007/978-3-319-93785-4_50.
2. Wigmans, R. *Calorimetry: Energy Measurement in Particle Physics*. (Oxford University Press, 2000). doi:10.1093/oso/9780198786351.001.0001.
3. Knoll, G. *Radiation Detection and Measurement*. (Wiley, 2010).
4. Pradeep Kumar, K. A., Shanmugha Sundaram, G. A., Sharma, B. K., Venkatesh, S. & Thiruvengadathan, R. Advances in gamma radiation detection systems for emergency radiation monitoring. *Nuclear Engineering and Technology* **52**, 2151–2161 (2020).
5. Gupta, T. K. *Radiation, Ionization, and Detection in Nuclear Medicine*. (Springer Berlin Heidelberg, Berlin, Heidelberg, 2013). doi:10.1007/978-3-642-34076-5.
6. Marques, L., Vale, A. & Vaz, P. State-of-the-Art Mobile Radiation Detection Systems for Different Scenarios. *Sensors* **21**, 1051 (2021).
7. Dhanekar, S. & Rangra, K. Wearable Dosimeters for Medical and Defence Applications: A State of the Art Review. *Adv Mater Technol* **6**, (2021).
8. Yang, Z., Vrielinck, H., Jacobsohn, L. G., Smet, P. F. & Poelman, D. Passive Dosimeters for Radiation Dosimetry: Materials, Mechanisms, and Applications. *Adv Funct Mater* **34**, (2024).
9. Posar, J. A., Petasecca, M. & Griffith, M. J. A review of printable, flexible and tissue equivalent materials for ionizing radiation detection. *Flexible and Printed Electronics* **6**, 043005 (2021).
10. Kržanović, N. *et al.* Performance Testing Of Selected Types of Electronic Personal Dosimeters in X- and Gamma Radiation Fields. *Health Phys* **113**, 252–261 (2017).

11. Kržanović, N., Stanković, K., Živanović, M., Đaletić, M. & Ciraj-Bjelac, O. Development and testing of a low cost radiation protection instrument based on an energy compensated Geiger-Müller tube. *Radiation Physics and Chemistry* **164**, 108358 (2019).
12. Shimaoka, T., Koizumi, S., J. H. & Kaneko. Recent progress in diamond radiation detectors. *Functional Diamond* **1**, 205–220 (2021).
13. Straume, T. *et al.* Compact Tissue-equivalent Proportional Counter for Deep Space Human Missions. *Health Phys* **109**, 277–283 (2015).
14. *Solid-State Radiation Detectors*. (CRC Press, 2017). doi:10.1201/b18172.
15. Nikl, M. & Yoshikawa, A. Recent R&D Trends in Inorganic Single-Crystal Scintillator Materials for Radiation Detection. *Adv Opt Mater* **3**, 463–481 (2015).
16. Magalotti, D., Placidi, P., Dionigi, M., Scorzoni, A. & Servoli, L. Experimental Characterization of a Personal Wireless Sensor Network for the Medical X-Ray Dosimetry. *IEEE Trans Instrum Meas* **65**, 2002–2011 (2016).
17. Archambault, L. *et al.* Plastic scintillation dosimetry: Optimal selection of scintillating fibers and scintillators. *Med Phys* **32**, 2271–2278 (2005).
18. Bartesaghi, G. *et al.* A real time scintillating fiber dosimeter for gamma and neutron monitoring on radiotherapy accelerators. *Nucl Instrum Methods Phys Res A* **572**, 228–230 (2007).
19. Antonello, M. *et al.* Tests of a dual-readout fiber calorimeter with SiPM light sensors. *Nucl Instrum Methods Phys Res A* **899**, 52–64 (2018).
20. Mazziotta, M. N. *et al.* A light tracker based on scintillating fibers with SiPM readout. *Nucl Instrum Methods Phys Res A* **1039**, 167040 (2022).
21. Beischer, B. *et al.* A high-resolution scintillating fiber tracker with silicon photomultiplier array readout. *Nucl Instrum Methods Phys Res A* **622**, 542–554 (2010).
22. Lv, S. *et al.* Online Radiation Beam Tracking by Using Full-Inorganic Scintillating Fibers. *Adv Opt Mater* **12**, (2024).
23. Liu, P. Z. Y., Suchowerska, N., Abolfathi, P. & McKenzie, D. R. Real-time scintillation array dosimetry for radiotherapy: The advantages of photomultiplier detectors. *Med Phys* **39**, 1688–1695 (2012).
24. Libanori, A., Chen, G., Zhao, X., Zhou, Y. & Chen, J. Smart textiles for personalized healthcare. *Nat Electron* **5**, 142–156 (2022).
25. Shi, J. *et al.* Smart Textile-Integrated Microelectronic Systems for Wearable Applications. *Advanced Materials* **32**, (2020).
26. Zhang, T. *et al.* Ultraflexible Glassy Semiconductor Fibers for Thermal Sensing and Positioning. *ACS Appl Mater Interfaces* **11**, 2441–2447 (2019).
27. Bayindir, M. *et al.* Metal–insulator–semiconductor optoelectronic fibres. *Nature* **431**, 826–829 (2004).
28. Nguyen-Dang, T. *et al.* Multi-material micro-electromechanical fibers with bendable functional domains. *J Phys D Appl Phys* **50**, 144001 (2017).
29. Yan, W. *et al.* Single fibre enables acoustic fabrics via nanometre-scale vibrations. *Nature* **603**, 616–623 (2022).
30. Gumennik, A. *et al.* All-in-Fiber Chemical Sensing. *Advanced Materials* **24**, 6005–6009 (2012).
31. Yan, W. *et al.* Thermally drawn advanced functional fibers: New frontier of flexible electronics. *Materials Today* **35**, 168–194 (2020).
32. Loke, G., Yan, W., Khudiyev, T., Noel, G. & Fink, Y. Recent Progress and Perspectives of Thermally Drawn Multimaterial Fiber Electronics. *Advanced Materials* **32**, (2020).
33. Dinu, N. Silicon photomultipliers (SiPM). in *Photodetectors* 255–294 (Elsevier, 2016). doi:10.1016/B978-1-78242-445-1.00008-7.
34. Gundacker, S. & Heering, A. The silicon photomultiplier: fundamentals and applications of a modern solid-state photon detector. *Phys Med Biol* **65**, 17TR01 (2020).
35. Loke, G. *et al.* Digital electronics in fibres enable fabric-based machine-learning inference. *Nature Communications 2021 12:1* **12**, 1–9 (2021).

36. Guo, Y. *et al.* Polymer Composite with Carbon Nanofibers Aligned during Thermal Drawing as a Microelectrode for Chronic Neural Interfaces. *ACS Nano* **11**, 6574–6585 (2017).
37. Rein, M. *et al.* Diode fibres for fabric-based optical communications. *Nature* **560**, 214–218 (2018).
38. Qu, Y. *et al.* Superelastic Multimaterial Electronic and Photonic Fibers and Devices via Thermal Drawing. *Advanced Materials* **30**, (2018).
39. Gupta, N. *et al.* A single-fibre computer enables textile networks and distributed inference. *Nature* **639**, 79–86 (2025).
40. Temelkuran, B., Hart, S. D., Benoit, G., Joannopoulos, J. D. & Fink, Y. Wavelength-scalable hollow optical fibres with large photonic bandgaps for CO2 laser transmission. *Nature* **420**, 650–653 (2002).
41. Sinclair, L. E., Hanna, D. S., MacLeod, A. M. L. & Saull, P. R. B. Simulations of a Scintillator Compton Gamma Imager for Safety and Security. *IEEE Trans Nucl Sci* **56**, 1262–1268 (2009).
42. Marion, J. S. *et al.* Thermally Drawn Highly Conductive Fibers with Controlled Elasticity. *Advanced Materials* **34**, (2022).
43. Faustini, L. & Martini, G. Bend loss in single-mode fibers. *Journal of Lightwave Technology* **15**, 671–679 (1997).
44. Rawal, A., Saraswat, H. & Sibal, A. Tensile response of braided structures: a review. *Textile Research Journal* **85**, 2083–2096 (2015).
45. Brunnschweiler, D. The Structure and Tensile Properties of Braids. *Journal of the Textile Institute Transactions* **45**, T55–T77 (1954).
46. Hajagos, T. J., Liu, C., Cherepy, N. J. & Pei, Q. High-Z Sensitized Plastic Scintillators: A Review. *Advanced Materials* **30**, (2018).
47. Francis, K. *et al.* Performance of the first prototype of the CALICE scintillator strip electromagnetic calorimeter. *Nucl Instrum Methods Phys Res A* **763**, 278–289 (2014).
48. McNabb, R. *et al.* A tungsten/scintillating fiber electromagnetic calorimeter prototype for a high-rate muon experiment. *Nucl Instrum Methods Phys Res A* **602**, 396–402 (2009).
49. Zhao, J. *et al.* On Analog Silicon Photomultipliers in Standard 55-nm BCD Technology for LiDAR Applications. *IEEE Journal of Selected Topics in Quantum Electronics* **28**, 1–10 (2022).
50. Vetter, K. Multi-sensor radiation detection, imaging, and fusion. *Nucl Instrum Methods Phys Res A* **805**, 127–134 (2016).
51. Liu, S.-X. *et al.* Performance of real-time neutron/gamma discrimination methods. *Nuclear Science and Techniques* **34**, 8 (2023).
52. Liu, Q. *et al.* Image reconstruction using multi-energy system matrices with a scintillator-based gamma camera for nuclear security applications. *Applied Radiation and Isotopes* **163**, 109217 (2020).
53. Parajuli, R. K., Sakai, M., Parajuli, R. & Tashiro, M. Development and Applications of Compton Camera—A Review. *Sensors* **22**, 7374 (2022).
54. Carchon, R., Moeslinger, M., Bourva, L., Bass, C. & Zendel, M. Gamma radiation detectors for safeguards applications. *Nucl Instrum Methods Phys Res A* **579**, 380–383 (2007).
55. Vanhavere, F. & Van Hoey, O. Advances in personal dosimetry towards real-time dosimetry. *Radiat Meas* **158**, 106862 (2022).
56. Shwartz, B. A. Scintillation Detectors in Experiments on High Energy Physics. in 211–230 (2017). doi:10.1007/978-3-319-68465-9_13.
57. Kroupa, M. *et al.* A semiconductor radiation imaging pixel detector for space radiation dosimetry. *Life Sci Space Res (Amst)* **6**, 69–78 (2015).
58. Rosenfeld, A. B. Electronic dosimetry in radiation therapy. *Radiat Meas* **41**, S134–S153 (2006).
59. Enlow, E. & Abbaszadeh, S. State-of-the-art challenges and emerging technologies in radiation detection for nuclear medicine imaging: A review. *Front Phys* **11**, (2023).

# Acknowledgments

We acknowledge support by DTRA (Award No. HDTRA1-20-2-0002) Interaction of Ionizing Radiation with Matter (IIRM) University Research Alliance (URA).

# Author Contributions

N.G. designed the fibre structures, characterized the materials, and fabricated the fibres. N.G., H.Q., J.K., and I.K. designed the laser and radiation measurement setups. N.G. and H.Q. performed the experiments to characterize the response of the fibres to various laser and radiation sources. N.G. and H.Q. performed tensile and bending measurements. N.G. and H.Q. analysed the fibre response data. N.G. performed optical transport and fibre response simulations. Y.F., O.H., and A.D. supervised the research. N.G., H.Q., A.D., O.H., and Y.F. wrote and revised the manuscript with input from all authors.

# Ethics Declarations

The authors declare no competing interests.

# Additional Information

**Supplementary Information**
Supplementary Information is available for this paper.

**Corresponding Author**
Correspondence and requests for materials should be addressed to Yoel Fink (yoel@mit.edu).

# Supplementary Information

**Connection Scheme for Hydrodynamic Stabilization**

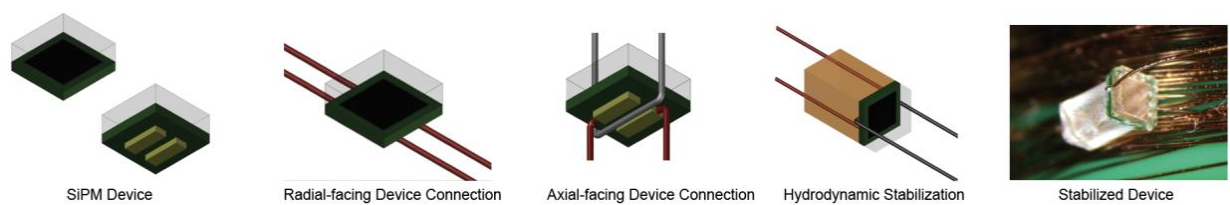

**Figure S1: Connection Scheme for Radial-to-Axial Conversion** | Schematic depicting the method by which the SiPM was connected for axial-facing integration into fibres. On the pad-face of the device, two 100-μm insulated copper wires (red) are electrically connected to the anode and cathode of the device. On the active photodetector array side, we loop a 100 μm insulated copper wire (grey) around the device. The wires are plastically deformed such that the photodetector array faces axially when an axial tension is imposed on the wires. Then, an epoxy domain is moulded on the pad-side of the device to make the axial dimension over double the planar dimension, causing the axial-facing configuration to now be more stable through the thermal draw process.

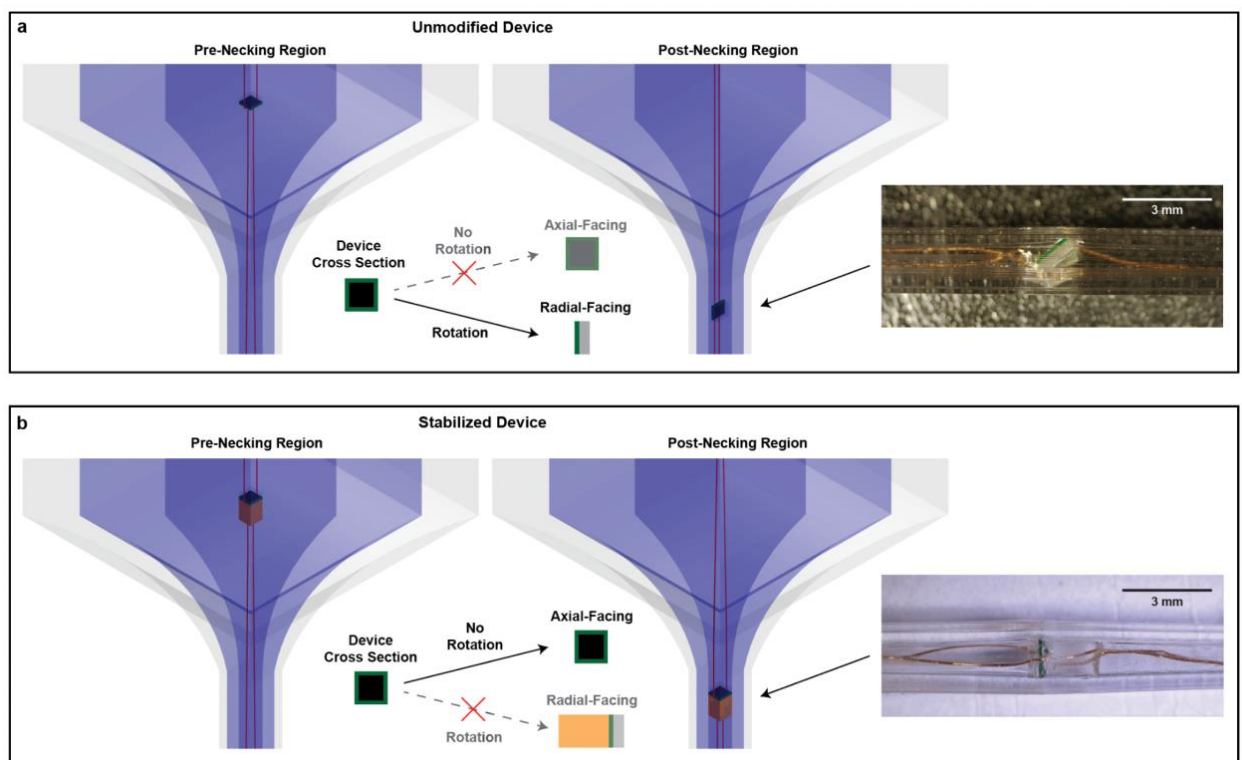

**Figure S2: Hydrodynamic Stabilization for Axial-Facing Device Integration** | Schematic depicting how the connection scheme in Figure S1 facilitates hydrodynamic stabilization of the device through the thermal draw process. **a.** When the device is drawn without the epoxy domain, the radial-facing configuration of the device presents the smallest hydrodynamic cross-section. Because of this, the device is prone to rotating during the thermal draw process to the more stable radial-facing configuration. **b.** With the epoxy domain moulded on the device, the axial-facing configuration now presents the smallest hydrodynamic cross-section. Therefore, the device now maintains the more stable axial-facing configuration through the thermal draw process.

**Plastic Scintillator Fibre Preform**

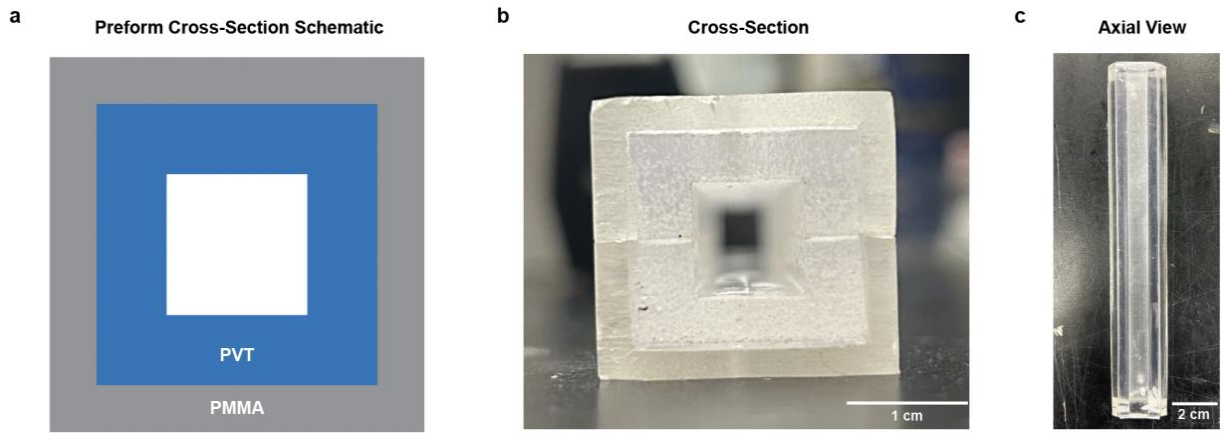

**Figure S3: Preform Architecture for the Plastic Scintillator Fibre | a.** A schematic of the preform cross section, which includes a rectangular annulus of PVT surrounded by a PMMA cladding. Images of the **b.** cross-sectional view and **c.** axial view of a consolidated preform.

**Rheological Properties of the Plastic Scintillator Fibre Materials**

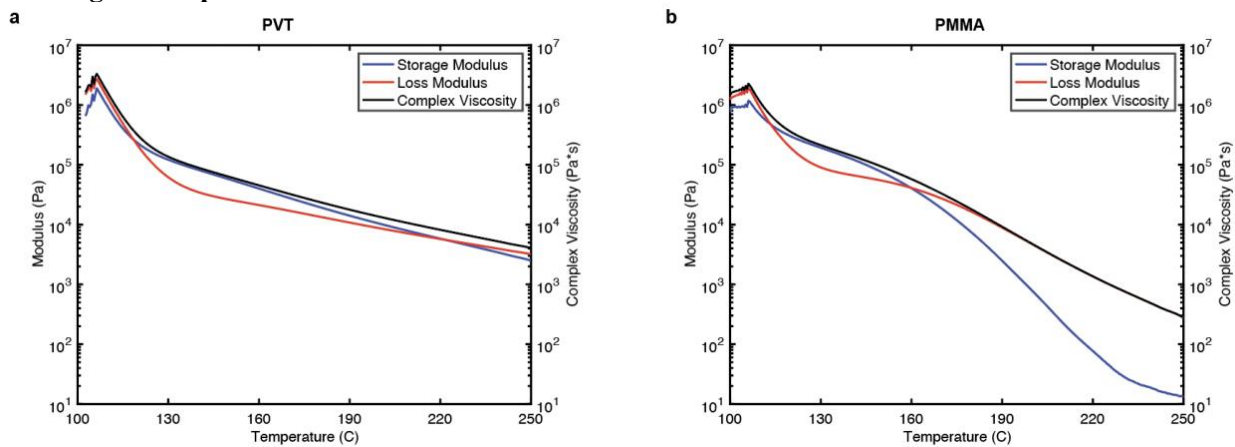

**Figure S4: Rheological Properties of the Plastic Scintillator Fibre Materials |** The storage modulus, loss modulus, and complex viscosity found from oscillatory shear rheological measurements over a temperature range of 100°C to 250°C for **a.** PVT and **b.** PMMA. Both materials maintain similar viscosities at draw temperatures, allowing for stable laminar co-flow of the two materials at high viscosity, which helps to preserve cross-sectional features through the draw. The storage modulus of PVT, which denotes its elastic response, has a large contribution at draw temperatures, so PMMA is used as an encapsulating cladding to maintain viscous thermal drawing conditions.

**Coupler Simulation**

The performance of the optical coupler in improving the coupling efficiency to the embedded SiPM in the plastic scintillator fibre was simulated using a 3D COMSOL ray optics simulation **(Figure S5)**. Rays were modelled to have a vacuum wavelength of 430 nm, which nearly corresponds to the wavelength of maximum emission of the PVT scintillator and the wavelength of peak photodetection efficiency of the SiPM. A 3 cm length of the plastic scintillator fibre was modelled, which includes a rectangular annulus of PVT (shown in blue), with a refractive index of 1.58, surrounded by a PMMA cladding (shown in grey) having a refractive index of 1.50. Although in reality the PMMA and PVT thicknesses thin out near the device, for simplicity, the dimensions of the fibre were chosen such that the air core of the fibre was exactly matched to the size of the SiPM, such that the fibre cross section is constant throughout. The SiPM (shown in green) was embedded in one end of the fibre, with dimensions corresponding to the manufacturer specifications and refractive index corresponding to that of the glass entrance window, which is specified to be 1.52. Next to the SiPM is the optical coupling region (shown in red), which fills a certain length of the air core of the fibre. The index of the coupler was set to be 1.576 based on the cured properties of the photopolymer used. At the other end of the fibre, rays were emitted from a point source (shown with a red dot) with a spherical emission pattern to represent a single excitation event in the fibre. Rays that reach the PMMA-air interface are frozen to only include rays that are guided in the PVT annulus. Through the simulation, the deposited ray power on the active area of the SiPM is measured until all the rays have fully propagated to the end of the fibre. The final transmittance is denoted to be the ratio between the total deposited ray power to the ray power emitted by the point source. Between simulations, we vary the coupler length to understand its effects on coupling efficiency to the embedded SiPM. The results show that transmittance increases with increasing coupler length up until 5 mm, at which point the coupling efficiency saturates **(Figure S5b)**, indicating that a coupler length of at least 5 mm is needed to achieve maximum coupling efficiency. We also simulated the effect of SiPM misalignment due to fabrication variability **(Figure S5c)**. The coupler length was fixed at 6 mm, while the SiPM was rotated at various angles from axial alignment **(Figure S5d)**, providing an estimate of optical performance variability attributable to device misalignment. As expected, coupling efficiency decreased with increasing angular deviation. However, even a 15° misalignment only reduced efficiency by less than 10%.

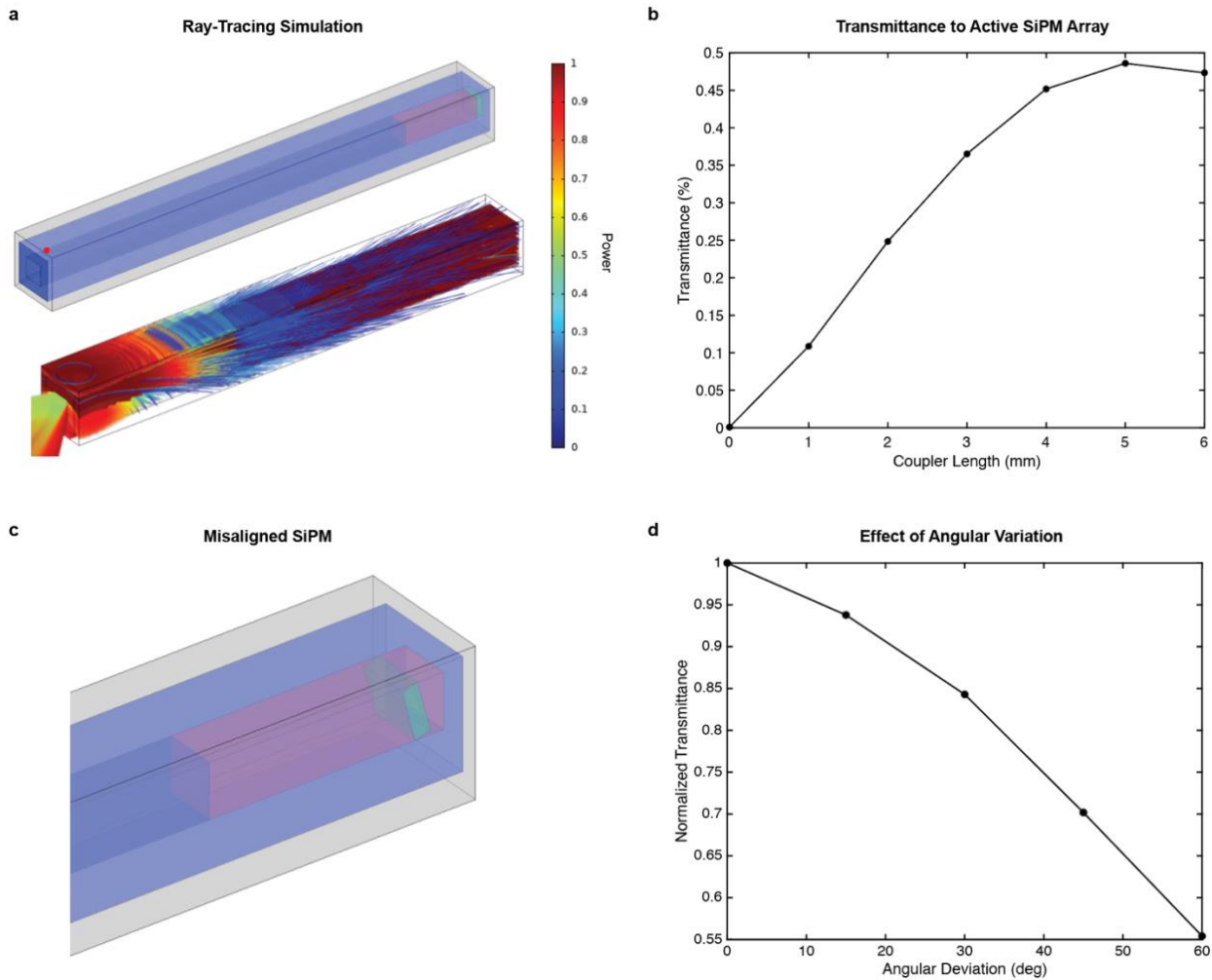

**Figure S5: Simulation of Optical Coupler | a.** A schematic of the COMSOL ray-tracing simulation used to model the plastic scintillator fibre, with the annular PVT waveguide (blue) surrounded by a PMMA (grey) cladding. At one of the fibre, a SiPM (green) is embedded in the air core, and is contacted by an optical coupling region (red) that fills a certain length of the air core. A spherical point emitter (red dot) was used to model a single excitation event 3 cm away from the embedded SiPM, from which rays propagated down the length of the fibre, with the colour bar indicating the power of each ray. **b.** The fraction of power emitted by the spherical point emitter that was collected by the active SiPM array as a function of the optical coupler length. **c.** A schematic of the simulation setup used to model the effect of SiPM misalignment relative to the fibre axis due to fabrication variability. The SiPM in the image is rotated 15° from axial alignment. **d.** The transmittance to the active SiPM array as a function of the angular deviation from axial alignment, normalized to the transmittance at perfect axial alignment.

## Single Photon Resolution

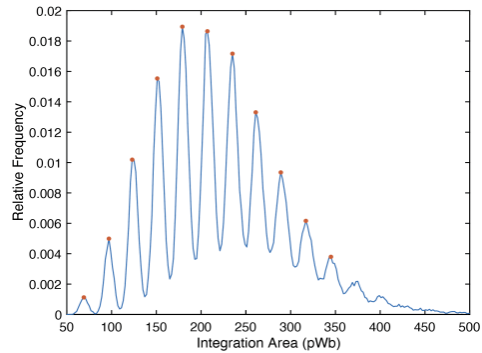

**Figure S6: Pulse Integration Spectrum** | Histogram of integrated pulse areas measured upon pulsed laser exposure of the plastic scintillator fibre. Each peak in the histogram corresponds to a discrete number of photons detected for each pulse of the laser. The red dots indicate the maxima of each photon peak, which were used to calibrate the integrated pulse area to the number of photons detected.

## Optical Measurements

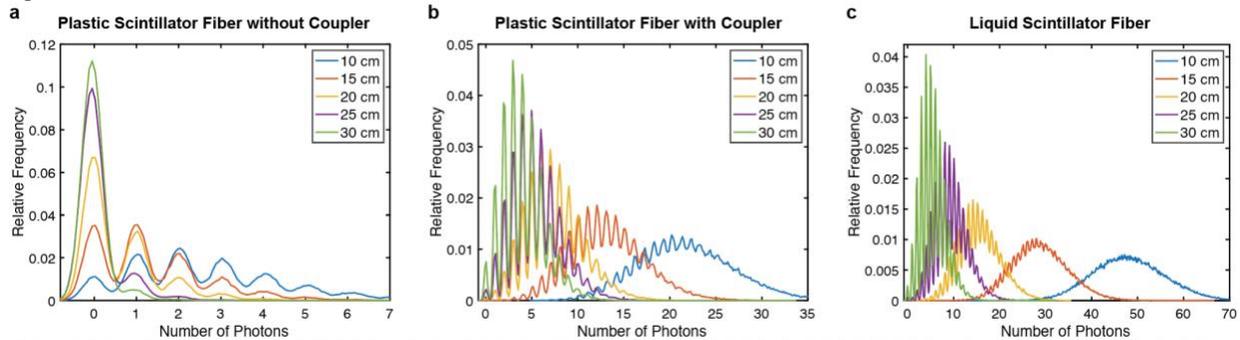

**Figure S7: Optical Measurements** | Photon spectra measured upon pulsed laser exposure at different distances of the laser away from the embedded SiPM for the **a.** plastic scintillator fibre without the optical coupler, **b.** plastic scintillator fibre with the optical coupler, and **c.** liquid scintillator fibre.

## Light-Blocking Layer

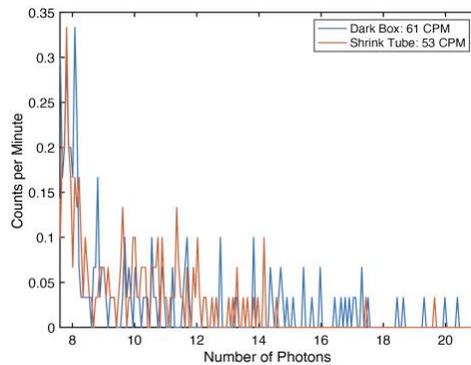

**Figure S8: Light-Blocking Layer** | Comparison of photon spectra measured for the plastic scintillator fibre while in a light-tight dark box, or while exposed to ambient lighting conditions when encapsulated in black polyolefin tubing that conforms to the fibre's profile. The figure legend shows the total photon count rate above the photon threshold for each of the conditions.

**Dark Count Noise**

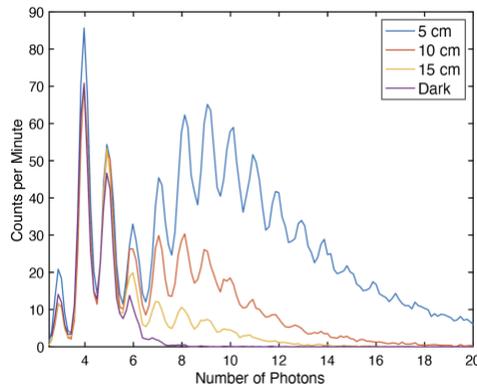

**Figure S9: Dark Count Noise** | Comparison of the photon spectrum in the absence of radiation (dark count noise) to examples of the photon spectra measured when Sr-90 radiation is incident on the plastic scintillator fibre at different distances away from the embedded SiPM.

**Radiation Measurements for the Liquid Scintillator Fibre**

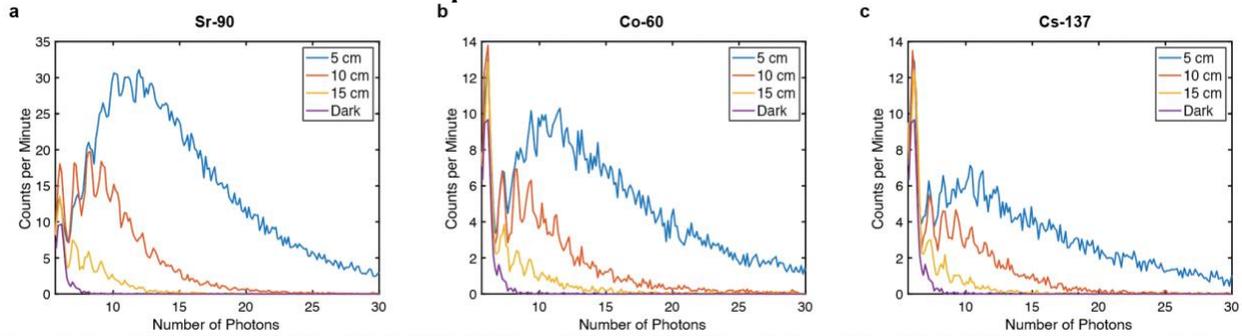

**Figure S10: Radiation Measurements for the Liquid Scintillator Fibre** | Photon spectra measured upon collimated radiation exposure to **a.** Sr-90, **b.** Co-60, and **c.** Cs-137 at different distances of the radiation source from the embedded SiPM.

**Radiation Measurements**

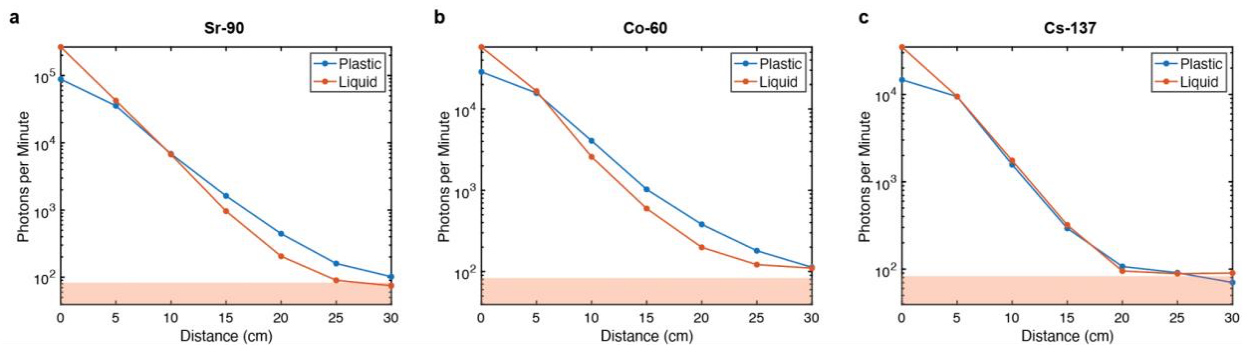

**Figure S11: Radiation Measurements** | Photon count rate above the photon threshold measured for each fibre in response to **a.** Sr-90, **b.** Co-60, and **c.** Cs-137. The red area in each plot indicates the dark count noise level in the absence of radiation.

**Modelling the Fibre Response to Arbitrary Fields**

Using the responsivity curves **(Figure 3b-d)**, we can model the total response of the fibres, $R_{tot}$, to arbitrary radiation fields by integrating the fibre's response to the local field intensity at each point along its length,

$$R_{tot} = \int_0^\infty R(z)I(z)dz \qquad (S1)$$

where $I(z)$ is the intensity of the radiation field at a certain axial distance $z$ away from the embedded SiPM. In the case of a uniform radiation field, this expression simplifies to,

$$R_{uniform} = I_{uniform} \int_0^\infty R(z)dz \qquad (S2)$$

where $I$ is the uniform intensity of the radiation field. In the case of a point source emitter, the expression becomes,

$$R_{tot} = \int_0^\infty R(z) * \left(\frac{A_{sim}}{4\pi(r^2 + (L-z)^2)}\right) dz \qquad (S3)$$

where $A_{sim}$ is the simulated point source activity, $r$ is the radial distance from the source to the fibre, and $L$ is the axial distance from the source to the embedded SiPM in the fibre. To simulate the response of the fibres in each of the scenarios, trapezoidal integration was performed in MATLAB according to the above expressions, with linear-log interpolation of the experimental responsivity measurements.

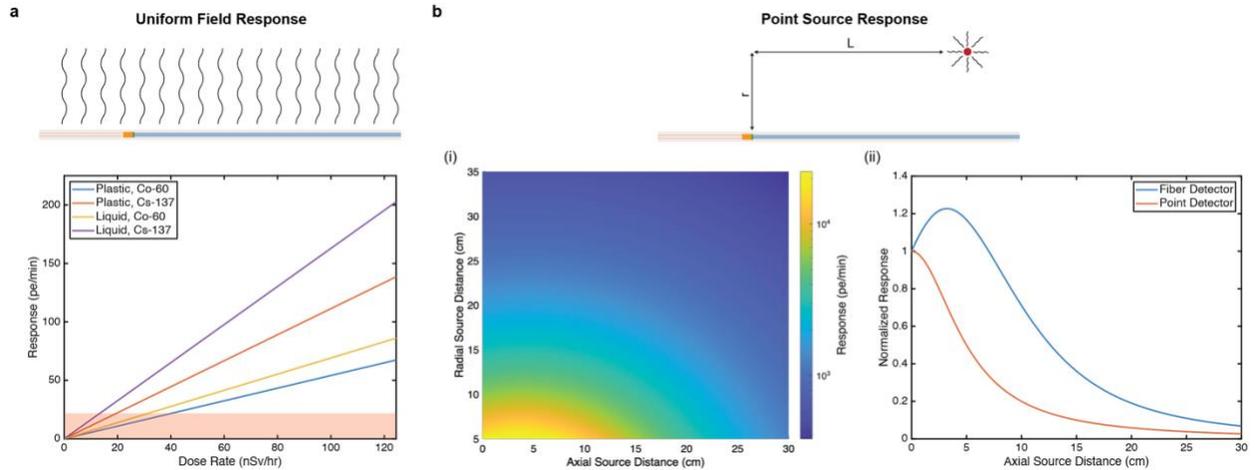

**Figure S12: Simulations of Fibre Response in Common Scenarios | a.** Simulated responses of the liquid and plastic scintillator fibres to uniform radiation fields. **b.** Simulated response of the plastic scintillator fibre to a Co-60 point source emitter. (i) Response of the fibre as a function of the radial and axial position of the source relative to the embedded SiPM. (ii) Comparison between the integrated fibre response and the inverse-square law response expected from a point detector, shown as a function of axial distance, $L$, for a constant radial source distance of $r = 5$ cm. The response of each type of detector is normalized to its response at $L = 0$ cm.

In **Figure S12a**, we show the simulated response of the fibres to uniform radiation fields according to Equation S2. Here, the gamma ray dose constants for Cs-137 and Co-60 (0.33 R/hr and 1.32 R/hr, respectively, for a 1 Ci source at 1 m), were used to convert field intensities to dose-rate units. Because the dose rate is independent of the position along the length of the fibre, the response scales linearly with the magnitude of the field. Although Co-60 has higher responsivity in terms of field intensity, Cs-137 has a significantly smaller gamma ray dose constant. As a result, when expressed in effective dose-rate units, the simulated response indicates that both fibres have higher sensitivity to Cs-137. To estimate a practical lower detection limit in the case of a uniform field, we determine the minimum dose rate that produces a count rate that is statistically distinguishable from the SiPM dark count noise in the absence of radiation. The experimentally measured variation in the dark count rate (as discussed in Methods) is indicated by the red area in Figure S12a. Therefore, the minimum detectable dose rate is therefore taken as the field at which the simulated fibre response is equal to the upper bound of the noise interval. Using this approach, the estimated lower detection limits for each fibre detector are summarized in Table S1. Because the responsivity near the embedded SiPM dominates the optical response in the uniform field case, the liquid scintillator fibre exhibits higher sensitivity due to its more favourable optical coupling geometry. Note that in mobile deployments, environmental background fluctuations may increase, which may raise the practical lower detection limit we can measure.

|  | Co-60 | Cs-137 |
| --- | --- | --- |
| Plastic Scintillator Fiber | 41 nSv/hr | 20 nSv/hr |
| Liquid Scintillator Fiber | 32 nSv/hr | 14 nSv/hr |

**Table S1: Lower Detection Limits** | The estimated Co-60 and Cs-137 lower detection limit for each fibre detector in a uniform field.

In **Figure S12b**, we use Equation S3 to show the simulated response of a plastic scintillator fibre to a Co-60 point source placed at various radial and axial distances away from the embedded SiPM. Because the waveguiding structure of the fibre allows for significant responsivity contributions over lengths approaching 30 cm, the fibre's response as the source is moved in the axial direction is higher than the inverse-square drop-off typically observed for radiation detectors that collect the response at a single location **(Figure S12b(ii))**. The extended sensitivity over the fibre's length is crucial for enabling the creation of distributed radiation detectors capable of capturing spatial variations in field intensity over large areas.